\documentclass[aps,prstab,preprint,groupedaddress,showpacs,showkeys,floatfix]{revtex4-1}

\usepackage{amssymb}

\usepackage[latin1]{inputenc}
\usepackage{graphicx}
\usepackage{color}
\usepackage{amsmath}
\usepackage{amssymb}
\usepackage{rotating}
\usepackage{epic}
\usepackage{placeins}
\usepackage{psfrag}
\usepackage{nicefrac}
\usepackage{cancel}
\usepackage{rotating}

\definecolor{mygray}{gray}{.5}
\newsavebox{\bbogen}
\sbox{\bbogen}{{%
\setlength\unitlength{1pt}%
\begin{picture}(4,3)
  \put(-1.0,-2.5){$\scriptscriptstyle\frown$}
  \put(-1.0,-1.0){$\scriptscriptstyle\frown$}
\end{picture}}}
\newsavebox{\bbogeng}
\sbox{\bbogeng}{{%
\setlength\unitlength{1pt}%
\begin{picture}(4,3)
  \put(-1.0,-2.5){$\scriptscriptstyle\color{mygray}{\frown}$}
  \put(-1.0,-1.0){$\scriptscriptstyle\color{mygray}{\frown}$}
\end{picture}}}
\newsavebox{\dbbogen}
\sbox{\dbbogen}{{%
\setlength\unitlength{1pt}%
\begin{picture}(4,3)
  \put(-1.5,-2.5){$\scriptscriptstyle\frown$}
  \put(-1.5,-1.0){$\scriptscriptstyle\frown$}
\end{picture}}}
\newsavebox{\bogen}
\sbox{\bogen}{{%
\setlength\unitlength{1pt}%
\begin{picture}(4,1)
  \put(-1.0,-2.5){$\scriptscriptstyle\frown$}
\end{picture}}}
\newsavebox{\bogeng}
\sbox{\bogeng}{{%
\setlength\unitlength{1pt}%
\begin{picture}(4,1)
  \put(-1.0,-2.5){$\scriptscriptstyle\color{mygray}{\frown}$}
\end{picture}}}
\newsavebox{\dbogen}
\sbox{\dbogen}{{%
\setlength\unitlength{1pt}%
\begin{picture}(4,1)
  \put(-1.5,-2.5){$\scriptscriptstyle\frown$}
\end{picture}}}

\newcommand{\flow}[1]{{%
  \setlength\unitlength{1pt}%
  \protect\begin{picture}(9,10)
    \put(0,0){\protect\makebox(9,10)[b]{%
       $\stackrel{\usebox{\bbogen}}{\protect\rule[0pt]{1pt}{0pt}#1}$}}
  \protect\end{picture}}}

\newcommand{\lowlowflow}[1]{{%
  \setlength\unitlength{1pt}%
  \protect\begin{picture}(9,10)
    \put(0,-2.5){\protect\makebox(9,10)[b]{%
       $\stackrel{\usebox{\bbogen}}{\protect\rule[0pt]{1pt}{0pt}#1}$}}
  \protect\end{picture}}}
\newcommand{\volt}[1]{{%
  \setlength\unitlength{1pt}%
  \protect\begin{picture}(9,9)
    \put(0,0){\protect\makebox(9,9)[b]{%
       $\stackrel{\usebox{\bogen}}{\protect\rule[0pt]{1pt}{0pt}#1}$}}
  \protect\end{picture}}}

\newlength{\widebarargwidth}
\newlength{\widebarargheight}
\newlength{\widebarargdepth}

\newcommand{\ve}{\volt{\bf e}}

\newcommand{\fb}{\flow{\bf b}}

\newcommand{\vh}{\volt{\bf h}}

\newcommand{\fd}{\flow{\bf d}}

\newcommand{\fj}{\lowlowflow{\bf j}}

\newcommand{\vel}{\hspace{0.5pt}\volt{\hspace{-0.5pt}e}}

\newcommand{\fbl}{\flow{b}}

\newcommand{\vhl}{\volt{h}}

\newcommand{\fdl}{\flow{d}}

\newcommand{\fjl}{\lowlowflow{j}}

\newcommand{\primC}{{\bf C}}
\newcommand{\dualC}{{\bf \widetilde{C}}}
\newcommand{\primS}{{\bf S}}
\newcommand{\dualS}{{\bf \widetilde{S}}}

\newcommand{\emit}{\varepsilon}

\def\pfeil{$\scriptscriptstyle\hspace{0.3em}\m@th\mathord\rightarrow$}
\def\newvec#1{\vbox{\m@th\ialign{##\crcr
      \hfil\pfeil\hfil\crcr\noalign{\kern\p@\nointerlineskip}
      $\hfil\displaystyle{#1}\hfil$\crcr}}}
\def\ppfeil{$\scriptscriptstyle\m@th\mathord\leftrightarrow$}
\def\tensor#1{\vbox{\m@th\ialign{##\crcr
      \hfil\ppfeil\hfil\crcr\noalign{\kern\p@\nointerlineskip}
      $\hfil\displaystyle{#1}\hfil$\crcr}}}
\def\ppfeilkursiv{$\scriptscriptstyle\hspace{0.3em}\m@th\mathord\leftrightarrow$}
\def\tensorkursiv#1{\vbox{\m@th\ialign{##\crcr
      \hfil\ppfeilkursiv\hfil\crcr\noalign{\kern\p@\nointerlineskip}
      $\hfil\displaystyle{#1}\hfil$\crcr}}}
\renewcommand{\grid}{\mathcal{G}}
\newcommand{\gridd}{\widetilde{\mathcal{G}}}
\newcommand{\fracpartialt}{\frac{\partial}{\partial t}}
\newcommand{\fractotalt}{\frac{\text{d}}{\text{d} t}}
\newcommand{\meq}{Maxwell's equations}
\newcommand{\Econt}{\vec{E}}
\newcommand{\Dcont}{\vec{D}}
\newcommand{\Bcont}{\vec{B}}
\newcommand{\Hcont}{\vec{H}}
\newcommand{\Jcont}{\vec{J}}

\newcommand{\Rhocont}{\rho}
\newcommand{\Fcont}{\vec{F}}
\newcommand{\vcont}{\vec{v}}
\newcommand{\pcont}{\vec{p}}
\newcommand{\dd}{\text{d}}
\newcommand{\ctilde}{\widetilde{c}}
\newcommand{\ampere}{Amp$\grave{\text{e}}$re}

\newcommand{\whitem}{\textcolor{white}{-}}
\newcommand{\Atilde}{\widetilde{A}}
\newcommand{\Vtilde}{\widetilde{V}}
\newcommand{\eg}{e.g.,~}
\newcommand{\ie}{i.e.,~}
\newcommand{\primed}[1]{#1^\prime}

\newcommand{\transp}{^\text{T}}

\renewcommand{\imath}{\text{i}\,}

\newcommand{\refseefig}[1]{(see Fig.~\ref{#1})}

\newcommand{\cm}{\text{~cm}}
\newcommand{\mm}{\text{~mm}}
\newcommand{\um}{\text{~$\mu$m}}

\newcommand{\milli}{\text{m}}

\newcommand{\eV}{\text{~eV}}

\definecolor{gray50}{rgb}{0.5,0.5,0.5}

\begin{document}


\title{Extension of the Finite Integration Technique including dynamic mesh
refinement and its application to self-consistent beam dynamics simulations}


\author{Sascha M. Schnepp}
\email[]{schnepp@gsc.tu-darmstadt.de}
\thanks{Author to whom any correspondence should be addressed.}
\affiliation{Graduate School of Computational Engineering, Technische Universit\"at 
Darmstadt, Dolivostr.~15, 64293~Darmstadt, Germany}
%
\author{Erion Gjonaj}
\email[]{gjonaj@temf.tu-darmstadt.de}
\author{Thomas Weiland}
\email[]{weiland@temf.tu-darmstadt.de}
\affiliation{Institut f\"ur Theorie Elektromagnetischer Felder, TEMF, 
Technische Universit\"at Darmstadt, Schlo\ss gartenstr.~8, 64289~Darmstadt, Germany}


\date{\today}

Draft version as submitted to Physical Review Special Topics: Accelerators and Beams.

\begin{abstract}
An extension of the framework of the Finite Integration Technique (FIT)
including dynamic and adaptive mesh refinement is presented. After recalling
the standard formulation of the FIT, the proposed mesh adaptation procedure is described.
Besides the linear interpolation approach,
a novel interpolation technique based on specialized spline functions for approximating the discrete electromagnetic
field solution during mesh adaptation is introduced.
The standard FIT on a fixed mesh and the new
adaptive approach are applied to a simulation test case with known analytical solution.
The numerical accuracy of the two methods are shown to be comparable.
The dynamic mesh approach is, however, much more efficient. This is also demonstrated for the
full scale modeling of the complete RF gun
at the Photo Injector Test Facility DESY Zeuthen (PITZ) on a single computer.
Results of a detailed design study addressing the effects of individual
components of the gun onto the beam emittance using a fully self-consistent
approach are presented.
\end{abstract}


\maketitle

\section{Introduction}

Memory consumption and CPU time represent the main limitations for large-scale
electromagnetic field computations. This is especially the case for accelerator
physics simulations involving self-consistent charged particle models based on the so-called
\textit{Particle-In-Cell} (PIC) method~\cite{Wolfheimer2006:NIM}. Simulations of this type
are an indispensable tool for the
design and optimization of particle accelerators since they offer
a full insight into the beam dynamics down to the particle level. This is especially important
for the simulation of low-energy sections of an accelerator,
where space charge forces heavily influence the beam but also for, \eg
dark current or electron cloud simulations.
Apart from the solution of Maxwell's equations on a discrete grid space,
PIC simulations also include the self-consistent solution of the equations
of motion. Thus, they provide a full description of the accelerator structure including
space-charge and wakefield effects~\cite{Birdsall1991}.

In this article we address the important problem of numerical efficiency
of such beam dynamics simulations using the Finite Integration Technique (FIT).
The FIT has been successfully applied for the simulation of a wide range
of electromagnetic problems in accelerator physics~\cite{Weiland1984,Gjonaj2006}.
We propose an extension of this method including dynamic mesh refinement
in order to locally adjust the spatial grid
resolution according to the dynamics of particles and fields. This leads to considerable savings
in the overall number of computational degrees of freedom and, thus, reduces the computational 
burden in PIC simulations.

The article is organized as follows.
After describing the governing equations in Sec.~\ref{sec:form-phys-probl},
a brief review of the FIT on static grids is given in Sec.~\ref{sec:finite-integr-fram},
which also serves for introducing the notation. Sec.~\ref{sec:extens-towards-trans}
describes the dynamic mesh refinement procedures and Sec.~\ref{sec:incl-charg-part}
addresses specific issues of charged particle simulations on dynamically
refined meshes. Sec.~\ref{sec:examples} contains two applications. First,
an example with 
known analytical solution is considered for investigating accuracy and efficiency
of the proposed method.
After, results of a detailed design study of the PITZ photo injector
using self-consistent PIC simulations are presented. The achievements
are summarized in Sec.~\ref{sec:conclusions}. 


\section{\label{sec:form-phys-probl}Formulation of the Physical Problem}

The electromagnetic part of the physical problem considered is described by \meq.
Their integral form reads
\begin{eqnarray}
  \label{eq:ampere}
  \int_{\partial A} \Hcont \cdot \dd \vec{s} & = & \whitem \int_A \left(\fracpartialt \Dcont + \Jcont\right) \cdot \dd \vec{A}, \\
  \label{eq:faraday}
  \int_{\partial A} \Econt \cdot \dd \vec{s} & = & - \int_A \fracpartialt \Bcont \cdot \dd \vec{A}, \\
  \label{eq:gaussel}
  \int_{\partial V} \Dcont \cdot \dd \vec{A} & = & \whitem \int_V \Rhocont \medspace\dd V, \\
  \label{eq:gaussmag}
  \int_{\partial V} \Bcont \cdot \dd \vec{A} & = & \whitem 0,
\end{eqnarray}
where $A$ and $V$ denote arbitrary surfaces and volumes. 
The magnetic field strength is indicated by $\Hcont$, $\Bcont$ denotes the magnetic flux density, $\Econt$ the electric
field strength, $\Dcont$ the dielectric flux density,
and $\Jcont$ the electric current density.
The electric and magnetic quantities are related
according to the constitutive equations
\begin{eqnarray}
  \label{eq:DeE}
  \Dcont & = & \epsilon\, \Econt,\\
  \label{eq:BuH}
  \Bcont & = & \mu\, \Hcont.
\end{eqnarray}
For the linear, isotropic, nondispersive materials considered here the permittivity $\epsilon$ and
the permeability $\mu$ are material dependent constants.
The conservation of charge follows from \ampere 's law~(\ref{eq:ampere}) and
Gauss' law~(\ref{eq:gaussel}) in the form of the continuity equation
\begin{equation}
  \label{eq:elchargeconservation}
  \fractotalt \int_V  \Rhocont \medspace \dd V + \int_{\partial V} \Jcont \cdot \dd \vec{A} = 0.
\end{equation}
It reflects that a temporal change of the total charge,
$Q = \int_V  \Rhocont \medspace \dd V$, contained in a volume is caused only
by a flow of electric current into or out of the volume.

The link between \meq\ and the motion of charged particles is
established by the Lorentz force
\begin{equation}
  \label{eq:lorentzforce}
  \Fcont = q \left( \Econt + \frac{\vec{u}^\prime}{\gamma} \times \Bcont \right),
\end{equation}
on the one hand and Newton's law
\begin{equation}
  \label{eq:newton2nd}
  \Fcont = \fractotalt \pcont,
\end{equation}
on the other hand.
Above, $q$ is the charge of the particle, $\gamma$ is the relativistic factor and 
$\vec{u}^\prime = \gamma\vec{v}$, with
$\vcont$ the velocity of the particle. The mechanical momentum is denoted by $\pcont$.
The equations of motion, thus, read
\begin{eqnarray}
  \label{eq:PDEpos}
  \fractotalt \vec{u} & = & \frac{\vec{u}^\prime}{\gamma}, \\
  \label{eq:PDEMom}
  \fractotalt \vec{u}^\prime & = & \frac{q}{m_0} \left(\Econt + \frac{\vec{u}^\prime}{\gamma} \times \Bcont \right),
\end{eqnarray}
where $\vec{u}$ is the position vector and $m_0$ the particle mass at rest.

\section{\label{sec:finite-integr-fram}The Finite Integration Framework}

\begin{figure}[bht]
  \centering
    \includegraphics[width=55mm]{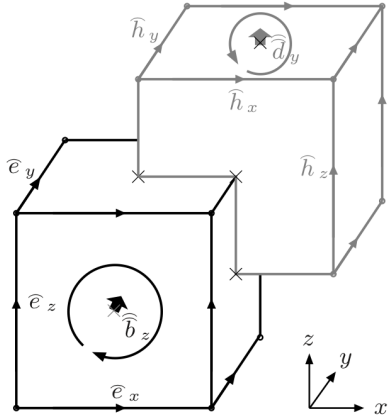}
    \caption{
      Staggered grid doublet in FIT. The primary grid and its associated quantities are depicted in black,
      the dual grid in gray.
    }
    \label{fig:GridDoublet}
\end{figure}

For introducing the notation as well as for completeness this section comprises a short 
review of the Finite Integration Technique.
For the spatial discretization of \meq, it utilizes
a staggered, dual orthogonal grid doublet $(\grid, \gridd)$ consisting of $N_p$ primary
nodes, which covers the domain of interest $\Omega$. 
The staggered grid arrangement is depicted in Fig.~\ref{fig:GridDoublet}.
Throughout this article, we employ the FIT on Cartesian grids
in three-dimensional space.
The following discrete integral state variables are introduced
\begin{equation}
  \label{eq:FITquantities}
  \begin{aligned}
    \vel & :=  \int_{c} \Econt \cdot \dd \vec{s}, \qquad
    &\vhl & :=  \int_{\ctilde} \Hcont \cdot  \dd \vec{s},\\
    \fbl & :=  \int_{A} \Bcont \cdot  \dd \vec{A}, \qquad
    &\fdl & :=  \int_{\Atilde} \Dcont \cdot \dd \vec{A},\\
    \fjl & :=  \int_{\Atilde} \Jcont \cdot \dd \vec{A}, \qquad
    &q    & :=  \int_{\Vtilde} \Rhocont\, \dd V,
  \end{aligned}
\end{equation}
where $c$ and $A$ belong to the set of edges and faces of the primary grid (black),
and $\ctilde, \Atilde$ and $\Vtilde$ belong to the set of edges, faces and volumes of its dual (gray).
The quantities $\vel$ and $\vhl$ represent grid voltages,
whereas $\fbl, \fdl$ and $\fjl$ are field fluxes on the faces of the grid.

Summing up the voltages $\vel$ and $\vhl$ along the four edges enclosing one face
yields a discrete but exact representation of \ampere 's and Faraday's law,
(\ref{eq:ampere}) and (\ref{eq:faraday}), in the form
\begin{eqnarray}
  \label{eq:FITampere}
  \sum\limits_{j=1}^4 \pm \vhl_j &=& \whitem \fractotalt \fdl + \fjl,\\
  \label{eq:FITfaraday}
  \sum\limits_{j=1}^4 \pm \vel_j &=& -\fractotalt \fbl,
\end{eqnarray}
for every face of the computational grid.
The orientation of the involved voltages $\vel_j, \vhl_j$ with respect to the orientation of the
loop integral in (\ref{eq:ampere}) and (\ref{eq:faraday}) determines the summation signs \refseefig{fig:GridDoublet}.

The discrete Gauss' laws are obtained in a similar way by summing up the
electric and magnetic flux variables $\fdl$ and $\fbl$ defined on all faces
of a cell. Fluxes pointing out of the cell are counted positive, incoming
fluxes negative. This yields the algebraic set of equations
\begin{eqnarray}
  \label{eq:FITgaussel}
  \sum\limits_{j=1}^6 \pm \fdl_j & = & q,\\
  \label{eq:FITgaussmag}
  \sum\limits_{j=1}^6 \pm \fbl_j & = & 0,
\end{eqnarray}
for every cell of $\grid$ and $\gridd$ corresponding to the Gauss' law
for electric and magnetic charges, respectively.

The constitutive equations~(\ref{eq:DeE}) and (\ref{eq:BuH})
translate to
\begin{eqnarray}
  \label{eq:FITconstitutive}
  \fdl & = & M_\epsilon \vel, \\
  \label{eq:FITconstitutive3}
  \fbl & = & M_\mu \vhl,
\end{eqnarray}
where $M_\epsilon$ and $M_\mu$ provide an appropriate mapping of grid voltages into fluxes~\cite{Weiland1984}.

Collecting the $N_p$ voltages and fluxes in the vectors $\ve, \vh, \fb, \fd$ and $\fj$ allows for stating
(\ref{eq:FITampere})-(\ref{eq:FITgaussmag}) for the complete computational domain in the compact form
\begin{eqnarray}
  \label{eq:FITfaradayMatrix}
  \primC \ve & = & -\fractotalt \fb, \\
  \label{eq:FITampereMatrix}
  \dualC \vh & = & \textcolor{white}{-} \fractotalt \fd + \fj,\\
  \label{eq:FITgausselMatrix}
  \dualS \fd & = & \mathbf{q}, \\
  \label{eq:FITgaussmagMatrix}
  \primS \fb & = & \mathbf{0}.
\end{eqnarray}
The summation signs are gathered together in 
the matrices $\primC, \dualC$ and $\primS, \dualS$. These matrices are, hence, purely topological.
They represent a discrete curl and divergence operator of the FIT.
It can be shown that these operators maintain the properties of their counterparts in continuum~\cite{Weiland1985}
in the sense that
\begin{eqnarray*}
  \primS \primC &= 0,& \quad \dualS \dualC = 0, \\
  \primC \dualS\transp &= 0,& \quad \dualC \primS\transp = 0.
\end{eqnarray*}
This allows, \eg for the derivation of a discrete continuity
equation [cf.~Eqn.~(\ref{eq:elchargeconservation})] in the form
\begin{equation}
  \label{eq:FITchargeconservationel}
  \fractotalt \mathbf{q} + \dualS \fj = \mathbf{0}.
\end{equation}

Eqns.~(\ref{eq:FITfaradayMatrix})-(\ref{eq:FITgaussmagMatrix}) are called
{Maxwell} grid equations (MGE). Typically,
the leap-frog scheme is applied for their integration in time.
Alternatively, one may apply the longitudinal-transverse split operator scheme
introduced in~\cite{Strang1968,Lau2005} for the time integration,
which offers better numerical dispersion properties.
The latter approach is referred to as LT-FIT standing for Longitudinal-Transverse-FIT.

\section{\label{sec:extens-towards-trans} Dynamic Mesh Refinement}

\begin{figure}[tbh]
  \centering
  \psfrag{a}[cc]{(a)}
  \psfrag{b}[cc]{(b)}
  \psfrag{e}[cc]{(c)}
  \psfrag{f}[cc]{(d)}
  \psfrag{c}[cb]{Base grid}
  \psfrag{d}[cb]{}
  \psfrag{PEC}[lc]{ }
  \psfrag{Level0}[cb]{\small{\textcolor{mygray}{Level 0}}}
  \psfrag{Level1}[cb]{\small{\textcolor{mygray}{Level 1}}}
  \psfrag{Level2}[cb]{\small{\textcolor{mygray}{Level 2}}}
  \includegraphics[width=0.9\columnwidth]{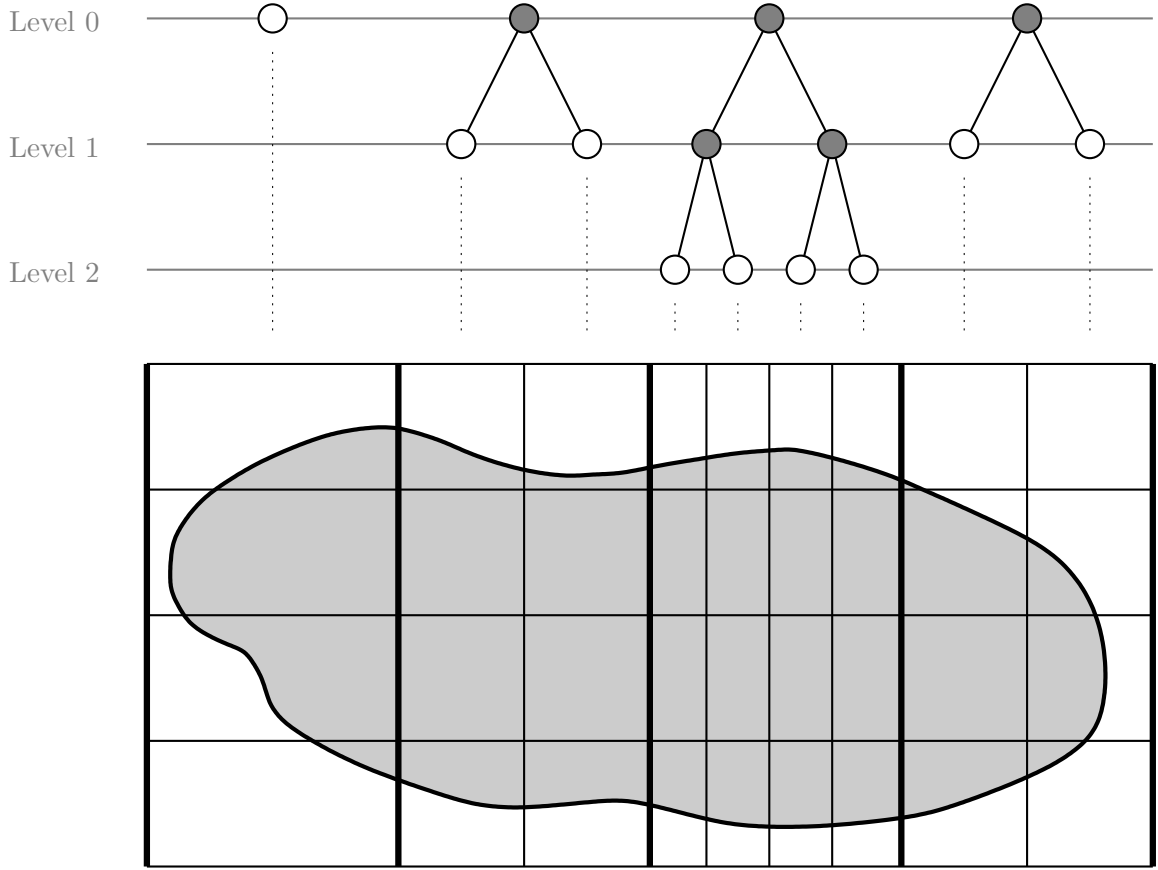}
  \caption{
    Representation of the mesh refinement using consecutive bisections and the refinement levels.
    Non-refined grid cells, such as the cells on the left end have refinement level zero.
    Each bisection increases the refinement level by one, which can be represented
    by means of a binary tree structure. The grid step size at level $L$ reduces as $1/2^L$.
  }
  \label{fig:treeStructure}
\end{figure}

We will restrict the discussion to conformal mesh refinements, which maintain the conformity and duality
of the Cartesian grid doublet. A refined mesh is obtained by a sequence of bisections applied
to a given mesh cell. The number of consecutive bisections
of a cell is referred to as its refinement level $L$ (see Fig.~\ref{fig:treeStructure}).

When mesh refinement is applied the discrete quantities assigned to the
primary and the dual grid need to be recomputed according to the modified mesh.
In Fig.~\ref{fig:AdapFITE} the refinement of a primary grid cell
and the subsequent arrangement of electric grid voltages is illustrated in
a two-dimensional cut view.
The preservation of the duality of the staggered grids requires adding
new nodes to the dual grid, as well as shifting existing nodes. This is
shown in Fig.~\ref{fig:AdapFITH}. In order to obtain a value for the new or shifted voltages
an interpolation procedure has to be carried out. Furthermore, it has to be distinguished
between the interpolation of voltages oriented in parallel to the refinement
[e.g.~$\primed{\vel}_x(\primed{2})$ in Fig.~\ref{fig:AdapFITE}(b)] and those
oriented perpendicularly [e.g.~$\primed{\vel}_z(2)$ and $\primed{\vel}_z(\primed{2})$
in Fig.~\ref{fig:AdapFITE}(c)]. In the following, two interpolation procedures,
linear and high order spline interpolation, will be discussed.

\subsection{Linear Interpolation of Grid Voltages}
\label{sec:adapFITinterpollinear}

\begin{figure}[htb]
    \vspace{3mm}
    \includegraphics[width=\columnwidth]{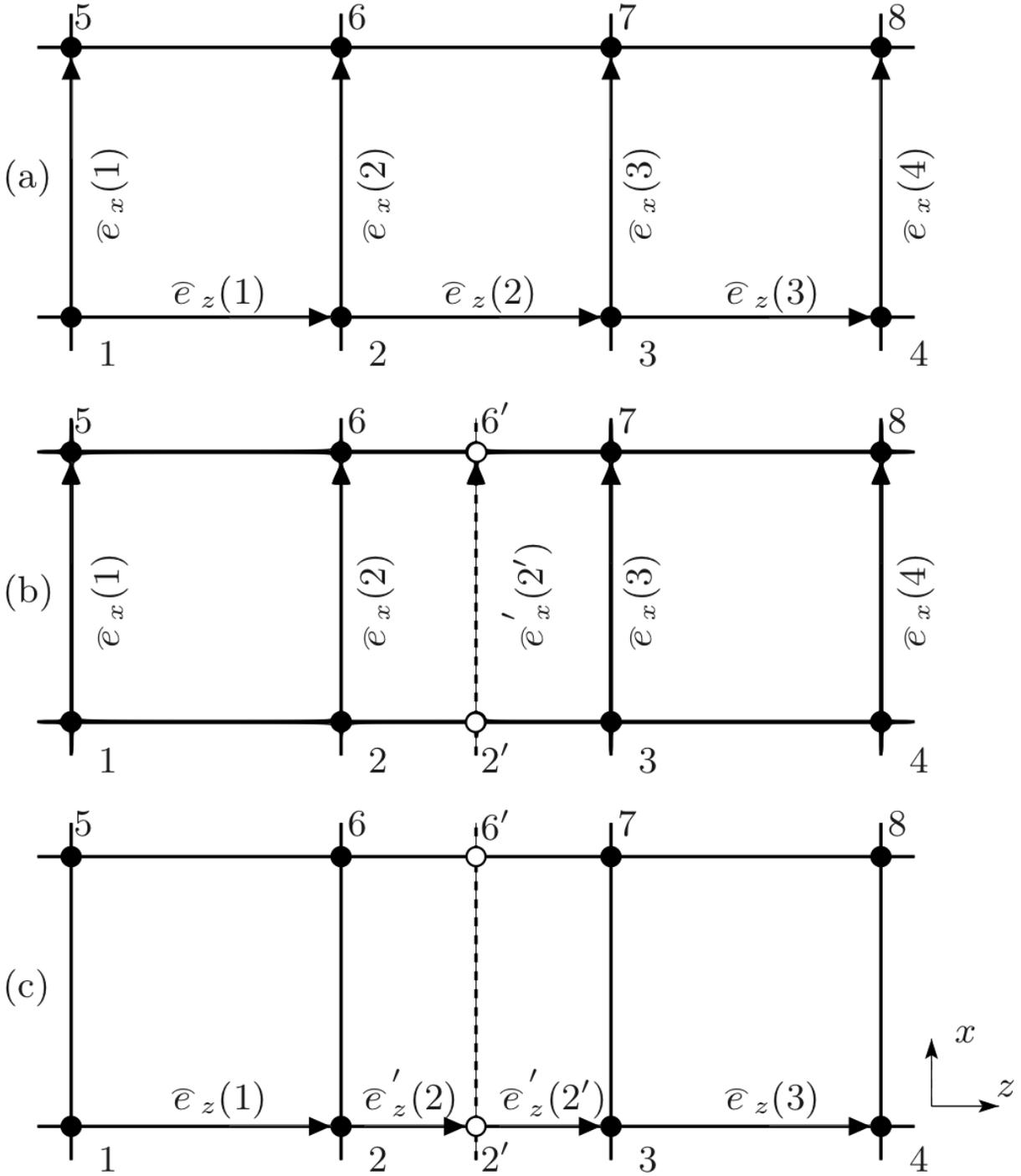}\\
  \caption{
      The refinement of the primary grid and
      the assignment of electric grid voltages to the refined edges is depicted in the
      $x-z$ plane. In (a) the initial situation is shown.
      In (b) and (c) the middle cell was refined, introducing
      an additional grid line (dashed). Thus, two
      nodes $\primed{2}$ and $\primed{6}$ (circles) are inserted.
      All voltages that are new or require an update of their value are marked
      with a prime. In (b) the position
      of the new electric grid voltage $\primed{\vel}_x(\primed{2})$ is shown.
      The voltage $\vel_z(2)$, oriented along the $z$-coordinate, has to be split into
      the two voltages $\primed{\vel}_z(2)$ and $\primed{\vel}_z(\primed{2})$ as shown in (c).\label{fig:AdapFITE}}
\end{figure}

\begin{figure}[htb]
    \psfrag{a}[cc]{}
    \psfrag{c}[cc]{}
    \psfrag{e}[cc]{}
    \psfrag{z1}[lb]{}
    \psfrag{z2}[lb]{}
    \psfrag{z3}[lb]{}
    \psfrag{z31}[lb]{}
    \psfrag{z7}[lb]{}
    \psfrag{z0}[lb]{}
    \psfrag{e1}[lc][1][1][90]{}
    \psfrag{e2}[lc][1][1][90]{}
    \psfrag{e3}[lc][1][1][90]{}
    \psfrag{e7}[lc][1][1][90]{}
    \psfrag{e0}[lc][1][1][90]{}
    \psfrag{P1}[lc]{}
    \psfrag{P2}[lc]{}
    \psfrag{P3}[lc]{}
    \psfrag{P4}[lc]{}
    \psfrag{P5}[lc]{}
    \psfrag{P6}[lc]{}
    \psfrag{P7}[lc]{}
    \psfrag{P8}[lc]{}
    \psfrag{P9}[lc]{}
    \psfrag{P0}[lc]{}
    \psfrag{x}[lb]{}
    \psfrag{z}[cb]{}
    \includegraphics[width=\columnwidth]{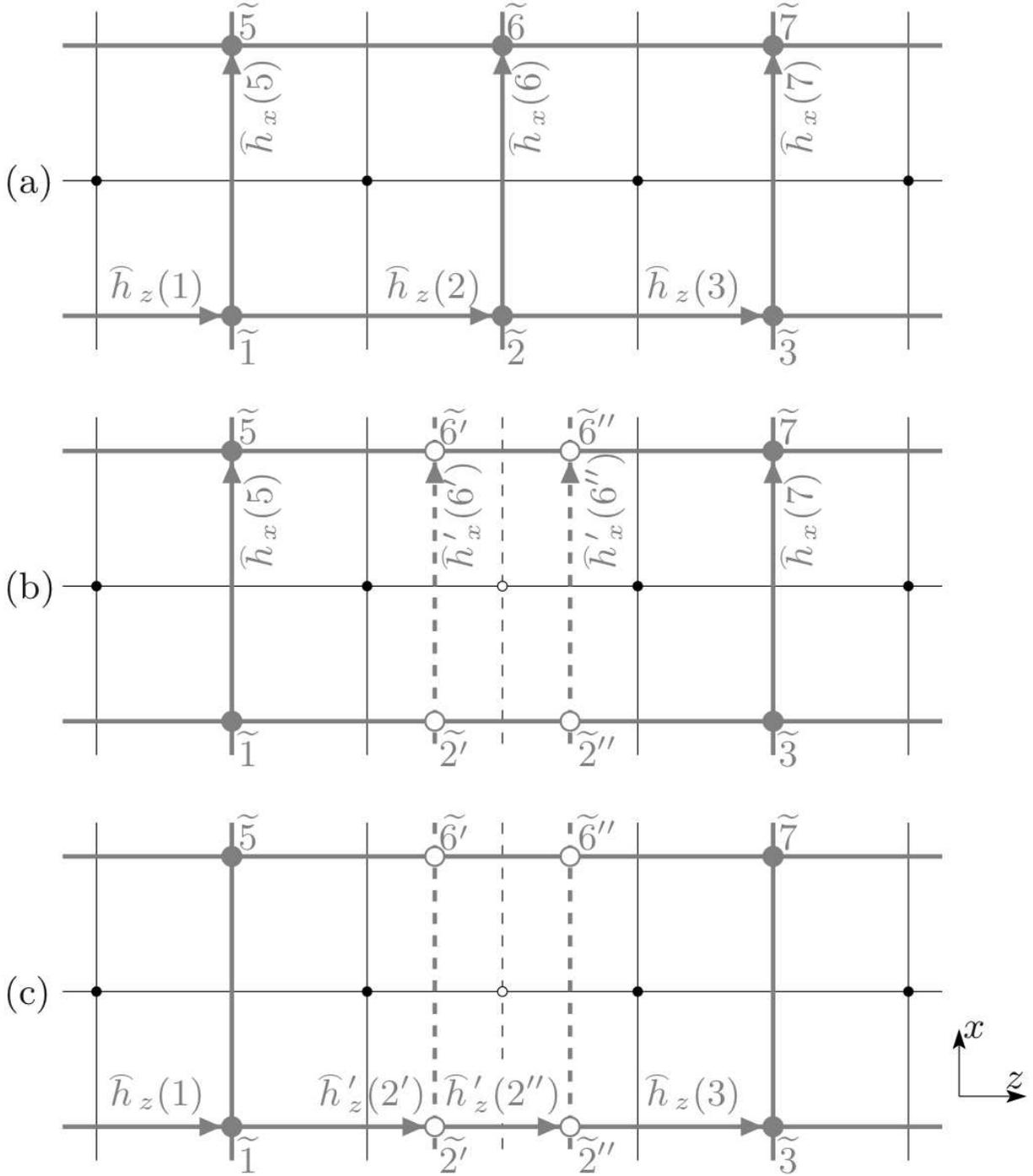}
  \caption{
      The refinement of the dual grid (gray), imposed
      by the refined primary grid (thin black lines) is depicted. The initial situation is shown
      in (a). In order to preserve the
      duality of the two grids the dual nodes $\widetilde{2}$ and $\widetilde{6}$
      have to be shifted. Their new position is indicated by $\widetilde{2}^\prime$ and $\widetilde{6}^\prime$
      in (b) and (c).
      In addition, the two new nodes $\widetilde{2}^{\prime\prime}$ and $\widetilde{6}^{\prime\prime}$ have to be
      inserted. In (b) the assignment of $x$-oriented magnetic voltages is shown and in (c)
      the assignment of $z$-oriented voltages.
      \label{fig:AdapFITH}}
\end{figure}

The application of a linear interpolation for the determination of
new or shifted voltages is straightforward.
The interpolation procedure described below refers to the refinement
scenario shown in Fig.~\ref{fig:AdapFITE} and \ref{fig:AdapFITH}.
The lengths of the primary and dual grid edges [cf.~\eqref{eq:FITquantities}] are denoted
by $|c|$ and $|\ctilde\,|$, respectively. 
Discrete field quantities associated with a specific node $i$ or
$\widetilde{i}$ are denoted as, \eg $c(i)$.
Primed quantities are either new or they have to be recomputed
during the adaptation procedure.

\paragraph{Refinement}

The values of new electric grid voltages oriented parallel to the refinement
[e.g.~$\primed{\vel}_x(\primed{2})$] are determined by
\begin{equation}
  \label{eq:FITetranslin}
  \primed{\vel}_x(\primed{2}) = \frac{\vel_x(2) + \vel_x(3)}{2}.
\end{equation}

\begin{figure}[tb]
    \psfrag{z1}[lb]{}
    \psfrag{z2}[lb]{}
    \psfrag{z3}[lb]{}
    \psfrag{z21}[lb]{}
    \psfrag{z22}[lb]{}
    \psfrag{a1}[lc]{}
    \psfrag{a0}[lc]{}
    \psfrag{a4}[lc]{}
    \psfrag{a2}[lc]{}
    \psfrag{a3}[lc]{}
    \psfrag{P1}[lc]{}
    \psfrag{P2}[lc]{}
    \psfrag{P3}[lc]{}
    \psfrag{P4}[lc]{}
    \psfrag{P5}[lc]{}
    \psfrag{z}[cc]{}
  \includegraphics[width=\columnwidth]{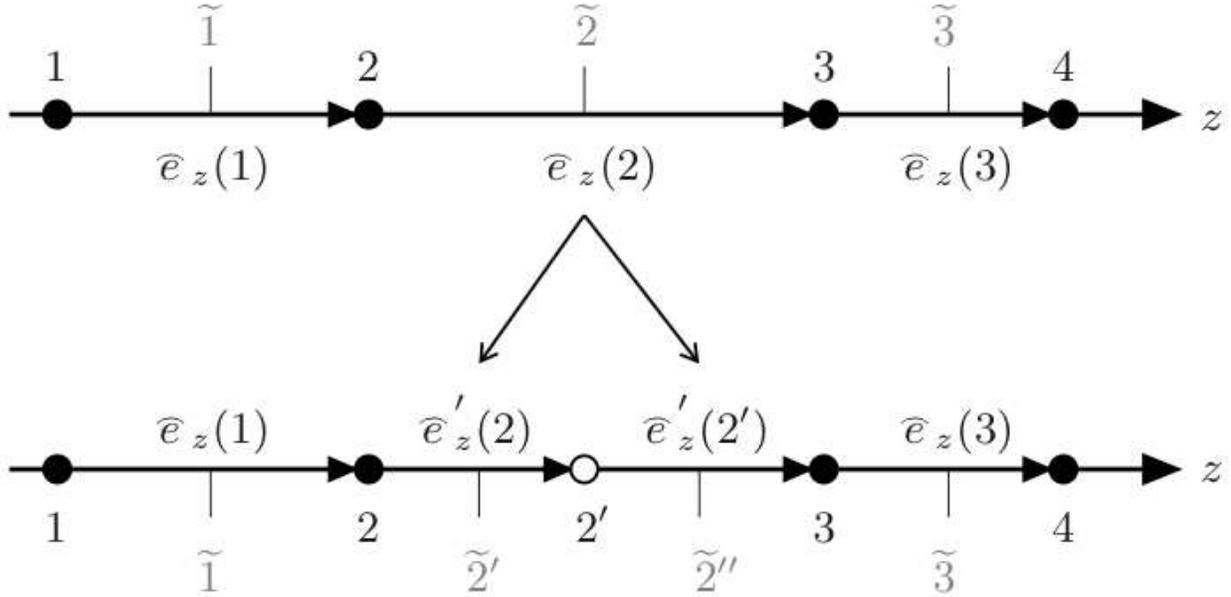}
  \caption{
      Interpolation of electric grid voltages oriented perpendicularly to the refinement.
      The coarse grid voltage $\vel_z(2)$ is split into the two voltages
      ${\vel^\prime_z(2)}$ and ${\vel^\prime_z(2^\prime)}$ such that the total voltage
      in between the nodes 2 and 3 is preserved [cf.~Eqn.~(\ref{eq:FITelonglin})].
      In order to assign values to ${\vel^\prime_z(2)}$ and ${\vel^\prime_z(2^\prime)}$
      the voltage gradient along the axis is evaluated using
      ${\vel_z(1)}$ and ${\vel_z(3)}$ [cf.~Eqns.~\eqref{eq:FITelongInter1}, \eqref{eq:FITelongInter2}].
      \label{fig:AdapFITEinterpol}}
\end{figure}

The splitting of voltages oriented perpendicularly to the refinement
[e.g.~$\vel_z(2)$] is constrained by the condition
\begin{equation}
  \label{eq:FITelonglin}
  \primed{\vel}_z(2) + \primed{\vel}_z(\primed{2}) \stackrel! = \vel_z(2).
\end{equation}
This condition imposes the conservation of the total voltage between the nodes 2 and 3.
A detailed sketch of this refinement is shown in
Fig.~\ref{fig:AdapFITEinterpol}.
In order to assign values to the refined voltages
${\vel^\prime_z(2)}$ and ${\vel^\prime_z(2^\prime)}$
the voltage gradient, \ie the average electric field, along the axis is approximated by
a central difference using the neighboring voltages
${\vel_z(1)}$ and ${\vel_z(3)}$.
The voltages assigned to the refined edges read
\begin{eqnarray}
  \label{eq:FITelongInter1}
  \primed{\vel}_z(2)          &=& |c^\prime(2)| \cdot \left( e_z(2) - \frac{|c^\prime(2)|}{2}
                                  \cdot \frac{\Delta e_z(3,1)}{ \Delta z(\widetilde{3},\widetilde{1})} \right),\\
  \label{eq:FITelongInter2}
  \primed{\vel}_z(\primed{2}) &=& |c^\prime(\primed{2})| \cdot \left( e_z(2) + \frac{|c^\prime(\primed{2})|}{2}
                                  \cdot \frac{\Delta e_z(3,1)}{\Delta z(\widetilde{3},\widetilde{1})} \right),
\end{eqnarray}
where the sampled electric field variables $e_z(i)$ are introduced as
\begin{equation}
  \label{eq:FITelongsample}
  e_z(i) := \frac{\vel_z(i)}{|c(i)|},
\end{equation}
the $z$-coordinate of the node $i$ is denoted by $z(i)$, and
the $\Delta$-notation denotes a difference of the respective quantity, \eg
$\Delta z(\widetilde{3},\widetilde{1}) = z(\widetilde{3}) - z(\widetilde{1})$.
Since cells are always divided into halves, $|c^\prime(2)|$ equals $|c^\prime(\primed{2})|$ and
the condition~\eqref{eq:FITelonglin} is fulfilled.

The interpolation of the magnetic voltages is more cumbersome. Besides the insertion
of new dual grid nodes, also existing nodes have to be shifted in position in order to
preserve the duality of the staggered grids. Shifting a node, however, implies
a modification of the grid voltages along the associated edges.

The interpolation of magnetic voltages oriented in parallel to the refinement is illustrated
in Fig.~\ref{fig:AdapFITH}(b). The voltages $\vhl^\prime_x(\widetilde{6}^{\prime})$
and $\vhl^\prime_x(\widetilde{6}^{\prime\prime})$
are given by
\begin{eqnarray}
  \label{eq:FIThtrans1}
  \vhl^\prime_x(\widetilde{6}^{\prime}) &=& \vhl_x(\widetilde{6}) - \Delta z(\widetilde{6}, \widetilde{6}^\prime)
  \cdot \frac{\Delta \vhl_x(\widetilde{7}, \widetilde{5})}{\Delta z(\widetilde{7}, \widetilde{5})}, \\
  \label{eq:FIThtrans2}
  \vhl^\prime_x(\widetilde{6}^{\prime\prime}) &=& \vhl_x(\widetilde{6}) 
  + \Delta z(\widetilde{6}^{\prime\prime}, \widetilde{6}) \cdot
  \frac{\Delta \vhl_x(\widetilde{7}, \widetilde{5})}{\Delta z(\widetilde{7}, \widetilde{5})}.
\end{eqnarray}

\begin{figure}[bt]
    \psfrag{e1}[lc]{}
    \psfrag{e2}[lc]{}
    \psfrag{e3}[lc]{}
    \psfrag{e4}[lc]{}
    \psfrag{e5}[lc]{}
    \psfrag{e6}[lc]{}
    \psfrag{e7}[lc]{}
    \psfrag{e8}[lc]{}
    \psfrag{e9}[lc]{}
    \psfrag{e0}[lc]{}
    \psfrag{z1}[cb]{}
    \psfrag{z2}[cb]{}
    \psfrag{z3}[cb]{}
    \psfrag{z4}[cb]{}
    \psfrag{z5}[cb]{}
    \psfrag{z6}[cb]{}
    \psfrag{z8}[cb]{}
    \psfrag{P1}[lc]{}
    \psfrag{P2}[lc]{}
    \psfrag{P3}[lc]{}
    \psfrag{P4}[lc]{}
    \psfrag{P5}[lc]{}
    \psfrag{P6}[lc]{}
    \psfrag{P7}[lc]{}
    \psfrag{P8}[lc]{}
    \psfrag{z}[cb]{}
    \includegraphics[width=\columnwidth]{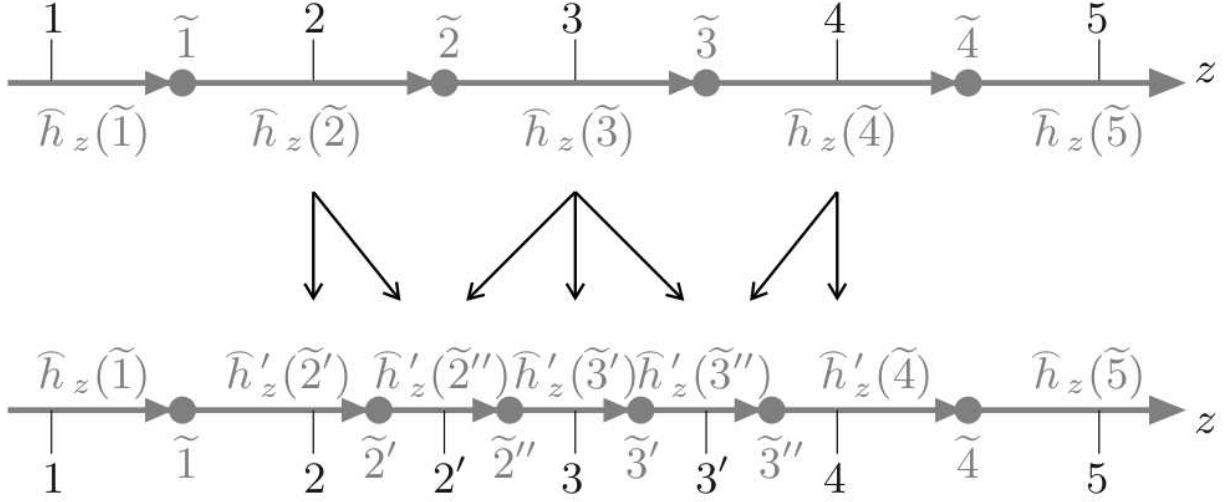}
  \caption{
      Interpolation of magnetic grid voltages oriented perpendicularly to the refinement.
      The voltages $\vhl_z(\widetilde{2}), \vhl_z(\widetilde{3})$ and $\vhl_z(\widetilde{4})$,
      assigned to coarse grid edges, are split
      and distributed such that the total magnetic voltage in
      between the nodes $\widetilde{2}$ and $\widetilde{4}$ is preserved
      [cf.~Eqn.~(\ref{eq:FIThlongCondition})]. The voltages on the refined grid are determined
      according to the [cf.~Eqns.~\eqref{eq:FIThlong1}-\eqref{eq:FIThlong5}].
      \label{fig:AdapFITHinterpol2}}
\end{figure}

Similarly to the splitting of electric voltages, the splitting of magnetic voltages, oriented perpendicularly
to the refinement, is constrained by the conservation of the total voltage.
However, for a complete description of the refinement algorithm,
the bisection of at least two neighboring cells has to be considered. This is illustrated in
Fig.~\ref{fig:AdapFITHinterpol2}. For this case, the conservation condition for the total voltage
between the nodes $\widetilde{1}$ and $\widetilde{4}$
can be expressed as
\begin{equation}
  \label{eq:FIThlongCondition}
  \vhl^\prime_z(\widetilde{2}^\prime) + \vhl^\prime_z(\widetilde{2}^{\prime\prime}) + \vhl^\prime_z(\widetilde{3}^\prime) + \vhl^\prime_z(\widetilde{3}^{\prime\prime})
  + \vhl^\prime_z(\widetilde{4})
  \stackrel! = \vhl_z(\widetilde{2}) + \vhl_z(\widetilde{3}) + \vhl_z(\widetilde{4}).
\end{equation}
The refined magnetic voltages read
\begin{align}
  \begin{split}
    \vhl^\prime_z(\widetilde{2}^\prime) =& \thickspace |\ctilde^{\,\prime}(\widetilde{2}^\prime)| \cdot \left[h_z(\widetilde{2})
      - \left( z(2) - \frac{z(\widetilde{1}) + z(\widetilde{2}^\prime) }{2}\right) \right. \\ 
    & \cdot \left. \frac{ \Delta h_z(\widetilde{4}, \widetilde{2}) }{ \Delta z(4, 2) } \right],
  \end{split}
  \label{eq:FIThlong1}\\ 
    \vhl^\prime_z(\widetilde{2}^{\prime\prime}) =& \thickspace \Delta 
    z(\widetilde{2}, \widetilde{2}^\prime) \cdot h_z(\widetilde{2}),
    + \Delta z(\widetilde{2}^{\prime\prime}, \widetilde{2}) \cdot  h_z(\widetilde{3}),
  \label{eq:FIThlong2}\\
    \vhl^\prime_z(\widetilde{3}^\prime) =& \thickspace |\ctilde^{\,\prime}(\widetilde{3}^\prime)| \cdot h_z(\widetilde{3}),
  \label{eq:FIThlong3}\\
    \vhl^\prime_z(\widetilde{3}^{\prime\prime}) =& \thickspace \Delta 
    z(\widetilde{3}, \widetilde{3}^\prime) \cdot h_z(\widetilde{3})
    + \Delta z(\widetilde{3}^{\prime\prime}, \widetilde{3}) \cdot  h_z(\widetilde{4}),
  \label{eq:FIThlong4}\\
  \begin{split}
    \vhl^\prime_z(\widetilde{4}) =&\thickspace |\ctilde^{\,\prime}(\widetilde{4})| \cdot  \left[h_z(\widetilde{4})
      + \left(\frac{z(\widetilde{3}^{\prime\prime}) + z(\widetilde{4}) }{2} - z(4)\right) \right. \\
    & \cdot \left. \frac{ \Delta h_z(\widetilde{4}, \widetilde{2}) }{ \Delta z(4, 2) } \right],
  \end{split}
  \label{eq:FIThlong5}
\end{align}
with the sampled magnetic field $h_z(\widetilde{i})$ defined as
\begin{equation}
  \label{eq:FIThsample}
  h_z(\widetilde{i}) := \frac{\vhl_z(\widetilde{i})}{|\ctilde(\widetilde{i})|}.
\end{equation}
Since the following equations hold true
\begin{equation}
  \label{eq:FITdualSum}
    z(2) - \frac{z(\widetilde{1}) + z(\widetilde{2}^\prime) }{2} = \frac{z(\widetilde{3}^{\prime\prime}) + z(\widetilde{4}) }{2} - z(4),
\end{equation}
and
\begin{eqnarray}
  \label{eq:FITdualSum2}
  |\ctilde(\widetilde{2})| &=& |\ctilde^{\,\prime}(\widetilde{2}^\prime)| + \Delta z(\widetilde{2},\widetilde{2}^\prime), \\
  \label{eq:FITdualSum3}
  |\ctilde(\widetilde{3})| &=& \Delta z(\widetilde{2}^{\prime\prime}, \widetilde{2}) 
  + |\ctilde^{\,\prime}(\widetilde{3}^\prime)| + \Delta z(\widetilde{3}, \widetilde{3}^\prime),\\
  \label{eq:FITdualSum4}
  |\ctilde(\widetilde{4})| &=& \Delta z(\widetilde{3}^{\prime\prime}, \widetilde{3}) + |\ctilde^{\,\prime}(\widetilde{4})|,
\end{eqnarray}
the Eqns.~\eqref{eq:FIThlong1}-\eqref{eq:FIThlong5} fulfill the condition~\eqref{eq:FIThlongCondition}.

The Eqns.~(\ref{eq:FITetranslin}), (\ref{eq:FITelongInter1}), (\ref{eq:FITelongInter2}),
(\ref{eq:FIThtrans1}), (\ref{eq:FIThtrans2}), and (\ref{eq:FIThlong1})-(\ref{eq:FIThlong5})
define the rules for performing linear interpolations of the
state variables $\ve$ and $\vh$. They enable consistent grid refinement based on cell bisection
within the FIT framework.

\paragraph{Coarsening}

In the following, the rules for performing grid coarsening are given. We refer again to the
Figs.~\ref{fig:AdapFITE} and \ref{fig:AdapFITH} and assume the removal of the
nodes $\primed{2}$ and $\primed{6}$ and the associated dual nodes.
Alongside with the removal of the primary nodes $\primed{2}$ and $\primed{6}$ and the connecting
edge, the electric grid voltage $\vel^\prime_x(\primed{2})$ is eliminated from the vector of
electric voltages.
No further modification is required for all electric voltages oriented in parallel to the refinement.
The constraint~\eqref{eq:FITelonglin} defines the rule for merging electric voltages
oriented perpendicularly to the refinement
\begin{equation}
  \label{eq:FITelongCoarsen}
  \vel_z(2) = \primed{\vel}_z(2) + \primed{\vel}_z(\primed{2}).
\end{equation}

For the determination of the magnetic coarse grid voltage $\vhl_x(\widetilde{6})$,
oriented parallel to the refinement, a linear interpolation of
the neighboring voltages allocated on the refined grid is performed
\begin{equation}
  \label{eq:FIThtransCoarsen}
  \vhl_x(\widetilde{6}) = \frac{ \primed{\vhl}_x(\widetilde{6}^\prime) + \primed{\vhl}_x(\widetilde{6}^{\prime\prime}) }{2}.
\end{equation}

The magnetic coarse grid voltages, oriented perpendicularly
to the refinement, are obtained by the
summation of fractional voltages given on the fine grid \refseefig{fig:AdapFITHinterpol2}.
The merging of the fine grid magnetic voltages
has to obey the constraint~\eqref{eq:FIThlongCondition}.
The coarse grid voltages are given by
\begin{align}
  \label{eq:FIThlongCoarsen1}
  \vhl_z(\widetilde{2}) =&\thickspace \primed{\vhl}_z(\widetilde{2}^\prime) +
  \Delta z(\widetilde{2}, \widetilde{2}^\prime)\cdot \primed{h}_z(\widetilde{2}^{\prime\prime}), \\
  \label{eq:FIThlongCoarsen2}
  \begin{split}
    \vhl_z(\widetilde{3}) =& \thickspace \Delta z(\widetilde{2}^{\prime\prime}, \widetilde{2})
    \cdot \primed{h}_z(\widetilde{2}^{\prime\prime})\\
    & + \primed{\vhl}_z(\widetilde{3}^\prime) 
    + \Delta z(\widetilde{3}, \widetilde{3}^\prime) \cdot \primed{h}_z(\widetilde{3}^{\prime\prime}),
  \end{split}\\
  \label{eq:FIThlongCoarsen3}
  \vhl_z(\widetilde{4}) =& \thickspace \Delta
  z(\widetilde{3^{\prime\prime}}, \widetilde{3}) \primed{h}_z(\widetilde{3}^{\prime\prime})
  + \primed{\vhl}_z(\widetilde{4}).
\end{align}
Since the following holds true
\begin{eqnarray}
  \label{eq:FITdualSum5}
  |\ctilde(\widetilde{2}^{\prime\prime})|
  &=& \Delta z(\widetilde{2}, \widetilde{2}^\prime) + \Delta z(\widetilde{2}^{\prime\prime}, \widetilde{2}),\\
  |\ctilde(\widetilde{3}^{\prime\prime})|
  &=& \Delta z(\widetilde{3}, \widetilde{3}^\prime) + \Delta z(\widetilde{3}^{\prime\prime}, \widetilde{3}),
\end{eqnarray}
the condition~\eqref{eq:FIThlongCondition} is fulfilled also for the case of grid coarsening.

Eqns.~\eqref{eq:FITetransCoarsen}-\eqref{eq:FIThlongCoarsen3} describe the
required modifications of the FIT field quantities during the coarsening process
of the computational grid. In combination with the refinement rules given above, they provide a
consistent framework for performing time-adaptive conformal grid refinement within the FIT.
Refinement schemes other than by cell bisecting are generally possible. However, these
are associated with a larger complexity in mesh administration and will not be
considered in this work.

\subsection{Spline Interpolation}
\label{sec:adapFITsplineInterpolation}

A linear interpolation of the discrete field quantities is easy to implement and 
fast in code execution. The interpolated quantities are second order accurate, which is
in agreement with the theoretical order of accuracy of the FIT. However, applying higher order
interpolating functions may be beneficial for obtaining a smoother representation of high-frequency fields
on the refined mesh.

Polynomials offer a possibility for performing higher order interpolations.
However, polynomials of high degrees tend to exhibit an oscillatory behavior and
possibly large overshoots, known as {Runge}{'s phenomenon}~\cite{Runge1901}.
Commonly, {spline functions} are employed in order to avoid this behavior.

\begin{figure}[tb]
    \psfrag{0}[cc]{\small 0}
    \psfrag{2}[cc]{\small 2}
    \psfrag{4}[cc]{\small 4}
    \psfrag{6}[cc]{\small 6}
    \psfrag{8}[cc]{\small 8}
    \psfrag{10}[cc]{\small 10}
    \psfrag{12}[cc]{\small 12}
    \psfrag{14}[cc]{\small 14}
    \psfrag{16}[cc]{\small 16}
    \psfrag{0,00}[cc]{\small 0.00}
    \psfrag{0,25}[cc]{\small 0.25}
    \psfrag{0,50}[cc]{\small 0.50}
    \psfrag{0,75}[cc]{\small 0.75}
    \psfrag{1,00}[cc]{\small 1.00}
    \psfrag{00,25}[cc]{\small -0.25}
    \psfrag{M1}[lc]{Data}
    \psfrag{M2}[lc]{Linear}
    \psfrag{M3}[lc]{C-Spline}
    \psfrag{M5}[lc]{Akima-Spline}
    \psfrag{x}[lc]{\Large $\frac{x}{\text{a.u.}}$}
    \psfrag{y}[cb]{\Large $\frac{f(x)}{\text{a.u.}}$}
    \includegraphics[width=0.95\columnwidth]{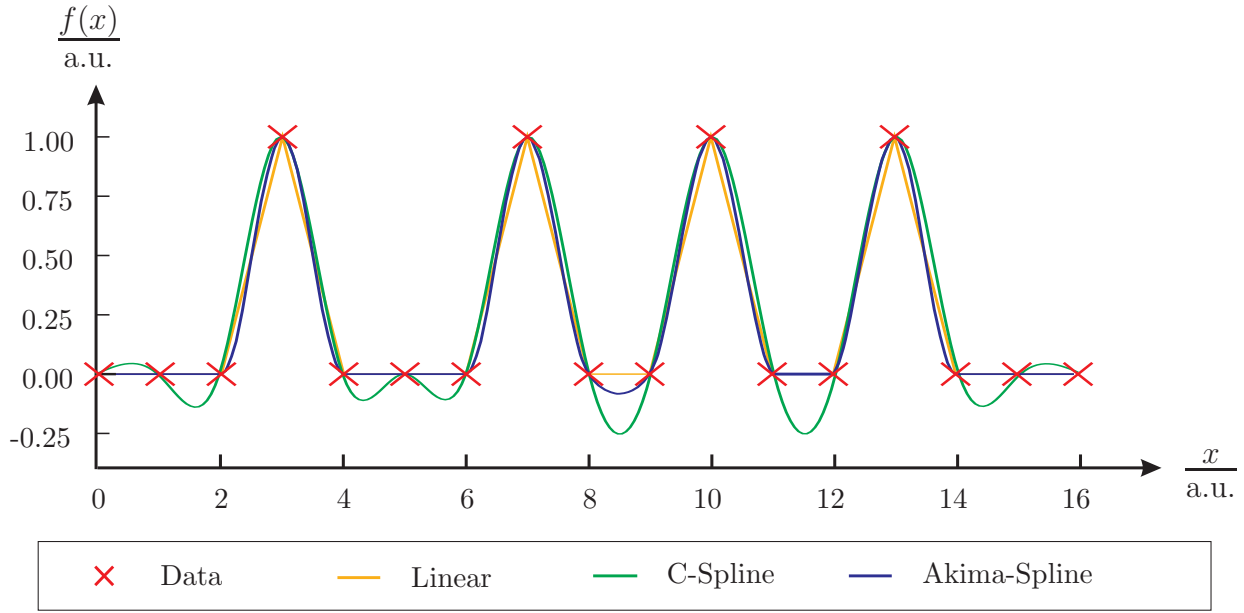}
  \caption{
    Comparison of spline interpolations. The given data points are interpolated using
    linear interpolation, a natural cubic spline (C-spline) and an {Akima}
    spline. The C-spline exhibits an oscillatory behavior
    and strong overshooting. Up to
    the data point 10, {Akima}'s original slope estimation was employed
    [cf.~Eqn.~(\ref{eq:splineSlope})].
    From this point on the minmod slope limiter [cf.~Eqn.~\eqref{eq:slopeMinmod}] was applied,
    which effectively avoids overshooting.
    \label{fig:AdapSplines}}
\end{figure}

A spline $S$ is a piecewise defined polynomial function of order $P$ which
is $P-1$ times continuously differentiable, i.e.,~$S \in \mathcal{C}^{P-1}$\cite{Boor2001}.
The construction rules determine the specific type of spline.
If the spline $S$ is demanded to pass exactly
through the given data points, and to be twice continuously
differentiable (i.e.~$S \in \mathcal{C}^2$) with the second derivative equal to zero on every interval
boundary, the so-called {natural cubic spline} (C-spline) is obtained. For its determination, a tridiagonal
system of equations has to be solved. During a time-domain simulation adopting time-adaptive grid
refinement, hundreds to thousands of grid adaptations are performed, each involving thousands of cells.
Thus, the solution of a system of equations
is computationally very expensive. In addition, the natural cubic spline may still exhibit overshoots
\refseefig{fig:AdapSplines}.

In order to further mitigate {Runge}'s phenomenon, the conditions on the
continuous differentiability of the spline have to be reduced. A spline $S$ of order
$P$ which is at most $P-2$ times continuously differentiable is called a {broken spline}
or {subspline}~\cite{Boor2001}. The {Akima} spline
is a cubic subspline which is an element of the functional space $\mathcal{C}^1$~\cite{Akima1970}. Cubic (sub-)splines
can conveniently be characterized by the triple of values $[x(i),f(i),s(i)]$ for each data point $i$,
where $x(i)$ is the coordinate of the data point, $f(i)$ its value, and $s(i)$ its slope.
Imposing the conditions
\begin{alignat*}{2}
  S[x(-1)] &= f(i-1), &\qquad\quad S[x(i)] &= f(i), \\
  \primed{S}[x(i-1)] &= s(i-1), &\qquad\quad \primed{S}[x(i)] &= s(i),
\end{alignat*}
at every interval boundary, uniquely describes the broken cubic spline
function
\begin{equation}
  \label{eq:splineFct}
  S(x) = a_0 + a_1 [x - x(i-1)] + a_2 [x - x(i-1)]^2 + a_3 [x - x(i-1)]^3,
\end{equation}
with the coefficients
\begin{eqnarray*}
  \label{eq:splineCoeffs}
  a_0 &=& f(i-1), \\
  a_1 &=& s(i-1), \\
  a_2 &=& \frac{3}{(\Delta x)^2} \Delta f(i, i-1) - \frac{1}{\Delta x} [s(i) + 2s(i-1)], \\
  a_3 &=& \frac{2}{(\Delta x)^3} \Delta f(i, i-1) + \frac{1}{(\Delta x)^2} \Delta s(i, i-1).
\end{eqnarray*}
However, the slopes $s(i)$ are, in general, unknown. The {Akima} 
procedure makes use of a local heuristic method for the estimation of the slopes.
It involves the data point $i$ and two neighboring points on each side.
First, the piecewise gradients $g$, given by
\begin{equation*}
  \label{eq:splineGradients}
  g(j) := \frac{f(j+1) - f(j)}{x(j+1) - x(j)}, \text{~~with~~} j = i-2, i-1, i, i+1,
\end{equation*}
are evaluated. These are weighted with the factors $w$
\begin{eqnarray*}
  \label{eq:splineWeights}
  w(i-1) &:=& |g(i+1) - g(i)|, \\
  w(i)   &:=& |g(i-1) - g(i-2)|,
\end{eqnarray*}
yielding the estimated slope at data point $i$
\begin{equation}
  \label{eq:splineSlope}
  s(i) := \frac{w(i-1) g(i-1) + w(i) g(i)}{w(i-1) +w(i)}.
\end{equation}
The fundamental idea of the heuristic can be summarized as:
the larger the difference between the two gradients on one side of point $i$, the larger
is the weighting factor applied to the gradient on the other side of this point.
This is intended to minimize overshooting and oscillatory behavior.
Note that, {Akima}'s heuristic is closely related to the idea of {slope limiters}.
These are commonly used in Finite Volume methods for reconstructing
a piecewise continuous solution on the grid~(cf.~\cite{VanLeer73I}).
Using, \eg the minmod limiter the estimated slope reads
\begin{multline}
  \label{eq:slopeMinmod}
  s(i) := \text{minmod}[g(i), g(i+1)] \\ =
  \small
  \begin{cases}
    \begin{array}{ll}
      g(i),     & \{\forall i \thickspace \big| \thickspace |g(i)| < |g(i+1)|, g(i)\, g(i+1) > 0\},\\
      g(i+1),   & \{\forall i \thickspace \big| \thickspace |g(i)| > |g(i+1)|, g(i)\, g(i+1) > 0\},\\
      0,        & \{\forall i \thickspace \big| \thickspace g(i)\, g(i+1) \le 0 \}.
    \end{array}
  \end{cases}
\end{multline}
Out of its two arguments, the minmod operator chooses the smaller one if they are equally signed
and zero otherwise. An overview of slope limiting techniques is found in \cite{LeVeque1990}.

In Fig.~\ref{fig:AdapSplines} the interpolation of a set of data points
using linear interpolation, the C-spline and the {Akima} spline is illustrated.
The C-spline shows an oscillatory behavior and strong overshooting.
For the determination of the {Akima} spline up to data point 10, the slope was calculated
using {Akima}'s slope definition~\eqref{eq:splineSlope}. For the points 11 to 16 the minmod
limiter was applied, which effectively avoids any overshooting.

\begin{figure}[tb]
    \psfrag{a}[cc]{(a)}
    \psfrag{c}[cc]{(b)}
    \psfrag{z1}[cb]{$1$}
    \psfrag{z2}[cb]{$2$}
    \psfrag{z3}[cb]{$3$}
    \psfrag{z4}[cb]{$4$}
    \psfrag{z5}[cb]{$5$}
    \psfrag{z6}[cb]{$6$}
    \psfrag{z7}[cb]{$7$}
    \psfrag{y2}[cb]{$\primed{2}$}
    \psfrag{y3}[cb]{$\primed{3}$}
    \psfrag{y4}[cb]{$\primed{4}$}
    \psfrag{y5}[cb]{$\primed{5}$}
    \psfrag{P1}[cc]{\textcolor{mygray}{$\widetilde{1}$}}
    \psfrag{P2}[cc]{\textcolor{mygray}{$\widetilde{2}$}}
    \psfrag{P5}[cc]{\textcolor{mygray}{$\widetilde{5}$}}
    \psfrag{P6}[cc]{\textcolor{mygray}{$\widetilde{6}$}}
    \psfrag{N2}[cc]{\textcolor{mygray}{$\widetilde{2}^\prime$}}
    \psfrag{O2}[cc]{\textcolor{mygray}{$\widetilde{2}^{\prime\prime}$}}
    \psfrag{N3}[cc]{\textcolor{mygray}{$\widetilde{3}^\prime$}}
    \psfrag{O3}[cc]{\textcolor{mygray}{$\widetilde{3}^{\prime\prime}$}}
    \psfrag{N4}[cc]{\textcolor{mygray}{$\widetilde{4}^\prime$}}
    \psfrag{O4}[cc]{\textcolor{mygray}{$\widetilde{4}^{\prime\prime}$}}
    \psfrag{N5}[cc]{\textcolor{mygray}{$\widetilde{5}^\prime$}}
    \psfrag{O5}[cc]{\textcolor{mygray}{$\widetilde{5}^{\prime\prime}$}}
    \psfrag{z}[lc]{$z$}
    \psfrag{y}[cc]{$S(z)$}
    \includegraphics[width=0.95\columnwidth]{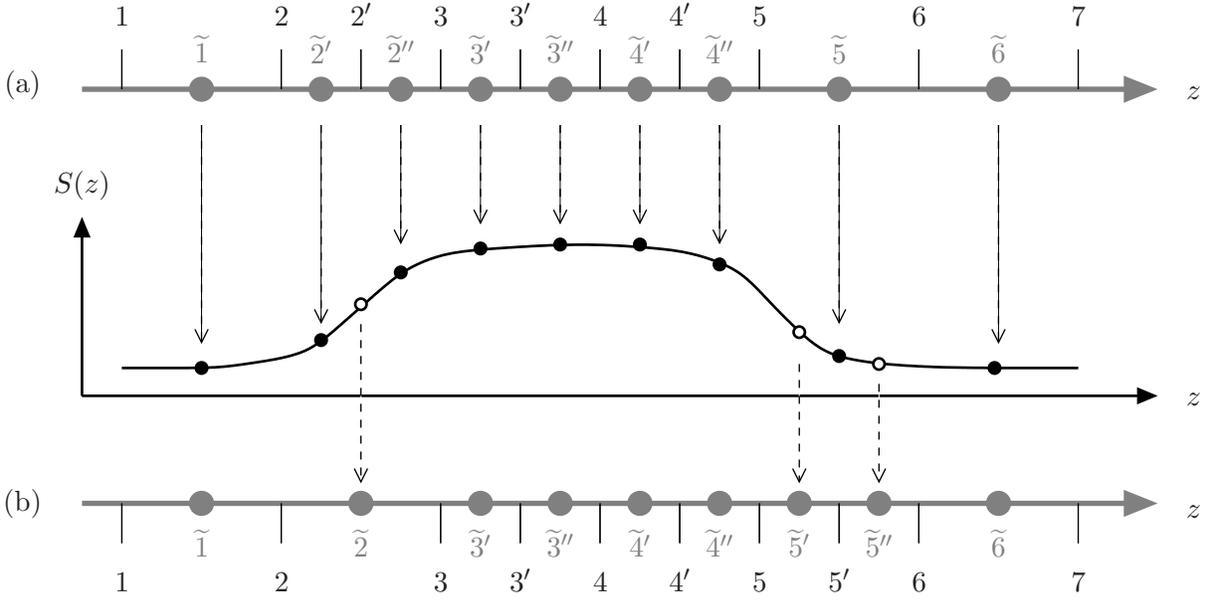}
  \caption{
    Setup and evaluation of the interpolating spline.
    An exemplary arrangement of grid nodes along the $z$-coordinate is shown
    in (a) and (b). In (b) the refined area of the grid has moved by a distance of one
    grid cell into the $z$-direction. Hence, the grid has to be coarsened around node 2
    and refined around node 5.
    The interpolating spline is set up, using the position of the nodes
    and field data in the refined area and some neighboring cells (black dots).
    An evaluation of the spline function at the position of new or shifted nodes
    (circles), yields the interpolated field values.\label{fig:AdapSplineInterpol}}
\end{figure}

The actual spline interpolation procedure of the discrete
field quantities is illustrated in Fig.~\ref{fig:AdapSplineInterpol}.
In (a) and (b) the primary and dual grid nodes along the $z$-coordinate
are indicated in black and gray. In (b) the refined
area has moved by a distance of one grid cell into the $z$-direction. Therefore, the grid topology
around the cells 2 (coarsening) and 5 (refinement) has to be modified.
The spline is set up using the position of the grid nodes
and the allocated field data in the refined area and some neighboring cells.
An evaluation of the spline function at the position of new or shifted nodes
yields the interpolated field values.

The {Akima} and the slope limited splines
are viable choices for performing interpolations within the computational grid.
They are formally third order accurate and can be fully constructed 
from local information. Thus, they do not require the
solution of a system of equations, which makes their implementation
very efficient for the purpose of adaptive grid refinement.

\section{\label{sec:incl-charg-part}Charged Particle Simulations on Adaptive Grids}

Simulations of the dynamics of charged particles using the FIT are performed routinely
and have first been reported in 1988 in~\cite{Bialowons1988}. For self-consistent simulations
the \textit{particle-in-cell} (PIC) algorithm is applied \cite{Birdsall1991a}. It
employs \textit{macro particles}, which may adopt any position from the space-continuous
domain of interest. Macro particles carry the charge and mass of about $10^{3..5}$ individual
particles. Due to~(\ref{eq:PDEMom}) the trajectory of a particle in the presence
of electromagnetic fields depends only on the ratio of its charge and rest mass, making
this a valid simplification.

The PIC algorithm consists of three steps. First, the electromagnetic field
at the space-continuous position of each macro particle has to be obtained from
the space-discrete FIT quantities. This is done by means of a trilinear interpolation.
Next, the equations of motion (\ref{eq:PDEpos}), (\ref{eq:PDEMom}) are integrated
using the algorithm described in~\cite{Boris1970}. Finally,
the convective currents are calculated from the position increment of
all particles. 
For this last step we adapted the cloud-in-cell (CIC) approach
\cite{Birdsall1968:cloud_in_cloud,Villasenor1992} to work with time-adaptive
mesh refinement. The particles are modeled as a uniformly charged volume of finite extent,
\ie a charged could.

\begin{figure}[tb]
    \centering
    \psfrag{t1}[cc]{$z$}
    \psfrag{t2}[cc]{$z$}
    \psfrag{t3}[cc]{$z$}
    \psfrag{t4}[cc]{$z$}
    \psfrag{t5}[cc]{$z$}
    \psfrag{t6}[cc]{$z$}
    \psfrag{t7}[cc]{$z$}
    \psfrag{t8}[cc]{$z$}
    \psfrag{t9}[cc]{$z$}
    \psfrag{1}[cc]{$1)$}
    \psfrag{2}[cc]{$2)$}
    \psfrag{3}[cc]{$3)$}
    \psfrag{4}[cc]{$4)$}
    \psfrag{5}[cc]{$5)$}
    \psfrag{6}[cc]{$6)$}
    \psfrag{7}[cc]{$7)$}
    \psfrag{8}[cc]{$8)$}
    \psfrag{9}[cc]{$9)$}
    \psfrag{P1}[cb]{2}
    \psfrag{P2}[cb]{3}
    \psfrag{P3}[cb]{$\primed{3}$}
    \psfrag{P4}[cb]{4}
    \psfrag{P5}[cb]{5}
    \psfrag{P6}[cb]{5}
    \psfrag{N1}[cc]{\textcolor{mygray}{$\widetilde{1}$}}
    \psfrag{N2}[cc]{\textcolor{mygray}{$\widetilde{2}$}}
    \psfrag{N3}[cc]{\textcolor{mygray}{$\widetilde{3}^\prime$}}
    \psfrag{N4}[cc]{\textcolor{mygray}{$\widetilde{3}^{\prime\prime}$}}
    \psfrag{N5}[cc]{\textcolor{mygray}{$\widetilde{4}$}}
    \psfrag{N6}[cc]{\textcolor{mygray}{$\widetilde{5}$}}
    \includegraphics[width=\columnwidth]{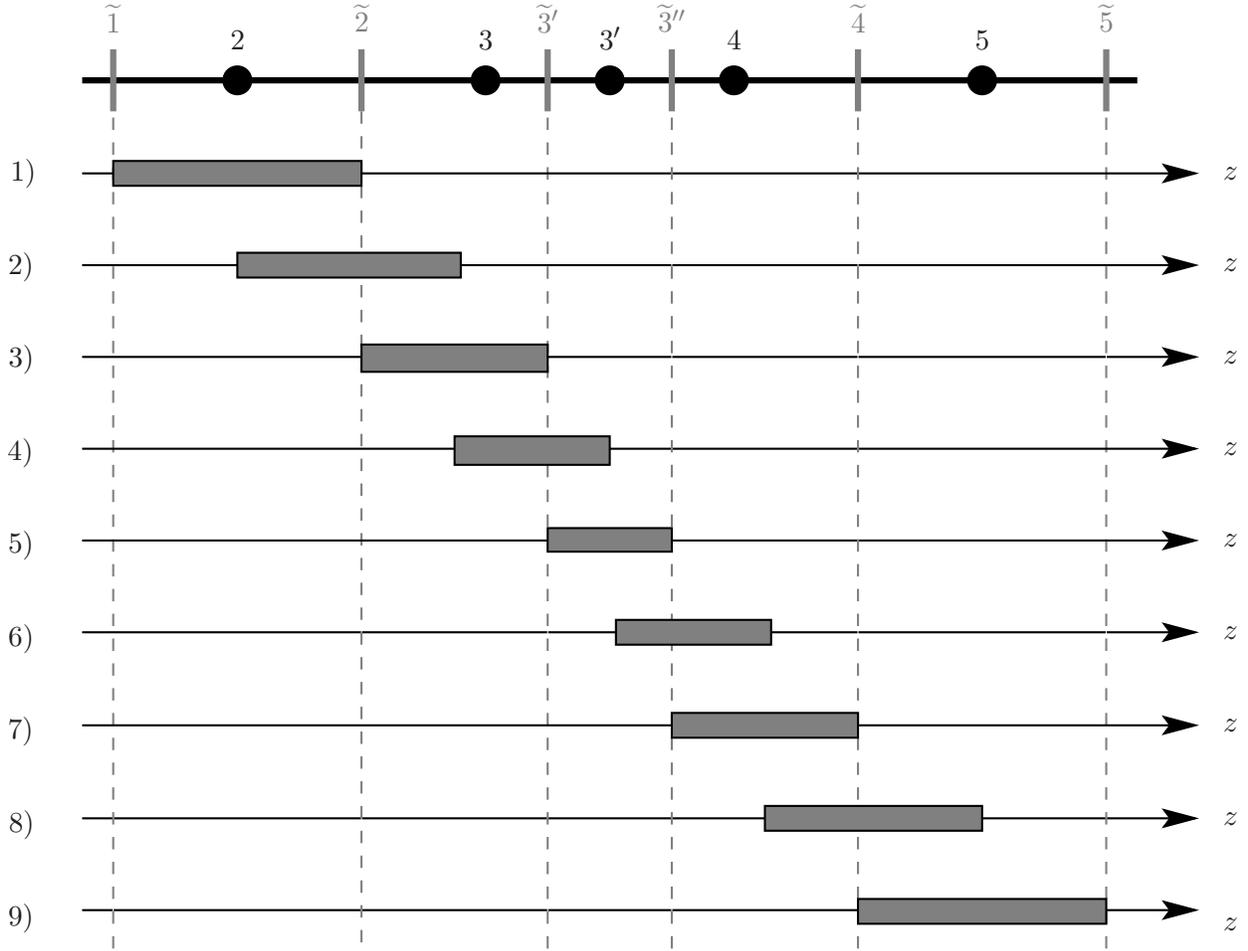}
  \caption{
    Illustration of the adaptation of the particle cloud size in one dimension.
    The size of the particle cloud relates to the extent of the covered cells on the dual grid.
    If the particle is centered within a dual grid cell (cases 1, 3, 5, 7 and 9) the sizes
    of the cloud and the dual volume coincide. Otherwise, the cloud has to be adapted
    asymmetrically around the particle position. For the cases 2 and 4 the charge density
    within the part of the cloud within the dual volumes $\widetilde{2}$
    or $\widetilde{3}^\prime$ has to be increased as a consequence of the diminished
    size. Contrarily, for the cases 6 and 8, the charge density of the cloud in the
    dual volumes $\widetilde{3}^{\prime\prime}$ and $\widetilde{4}$ has to be decreased.\label{fig:AdapCIC}}
\end{figure}

The extent of the cloud is usually chosen to coincide with the
size of the cells. However, if a nonequidistant grid is employed or in the case of adaptive
grid refinement, the nonuniform sizes of the cells demand for a modification of the
algorithm. 
There are two modification options: first, a constant size of the particle cloud is chosen
independently from the grid cell size or, second, the size of the cloud is adapted in order
to match the sizes of the involved cells. Depending on the local degree of grid refinement,
the first option can largely increase the number of cells affected by
one cloud. Besides the
coding efforts coming along with this non-local operation, the deposit of
fractional currents to a large number of grid points increases the computational
load. Therefore, the second option has been pursued. The adaptation of the
cloud in dependence of the grid cell size is illustrated in
Fig.~\ref{fig:AdapCIC} for a one-dimensional example.

The modified CIC approach maintains its charge conserving property, meaning that it fulfills
the discrete continuity equation~(\ref{eq:FITchargeconservationel}) in a cell-wise manner.


%

\section{Applications}
\label{sec:examples}

In this section, we present results of the application of the dynamic mesh adaptation
algorithm introduced above to the simulation of two setups.

\subsection{Bunch drift in a pipe}
\label{sec:pipe}

We consider a hollow, perfectly conducting pipe, which
is closed by a perfectly conducting plate at one end. The other end is left open. A bunch of
charged particles is emitted from the center of the end plate and travels along the pipe at a
constant velocity of $0.9\thinspace c_0$, where $c_0$ is the speed of light in vacuum.
The settings of the computational model are specified in Table~\ref{tab:tubeSettings}.
In~\cite{Onishchenko2002} the analytical solution of this problem for a semi-infinite pipe is given.
\begin{table*}[htb]
  \caption{\label{tab:tubeSettings}Settings of the benchmark example}
  \begin{ruledtabular}
    \begin{tabular}{lllll}
Pipe length  & Pipe radius & RMS bunch radius & RMS bunch length & Particle velocity \\
120 mm       & 40 mm       & 5 mm             & 3 mm                  & $0.9\, c_0$
    \end{tabular}
  \end{ruledtabular}
\end{table*}

\begin{figure*}[htb]
  \includegraphics[width=\textwidth]{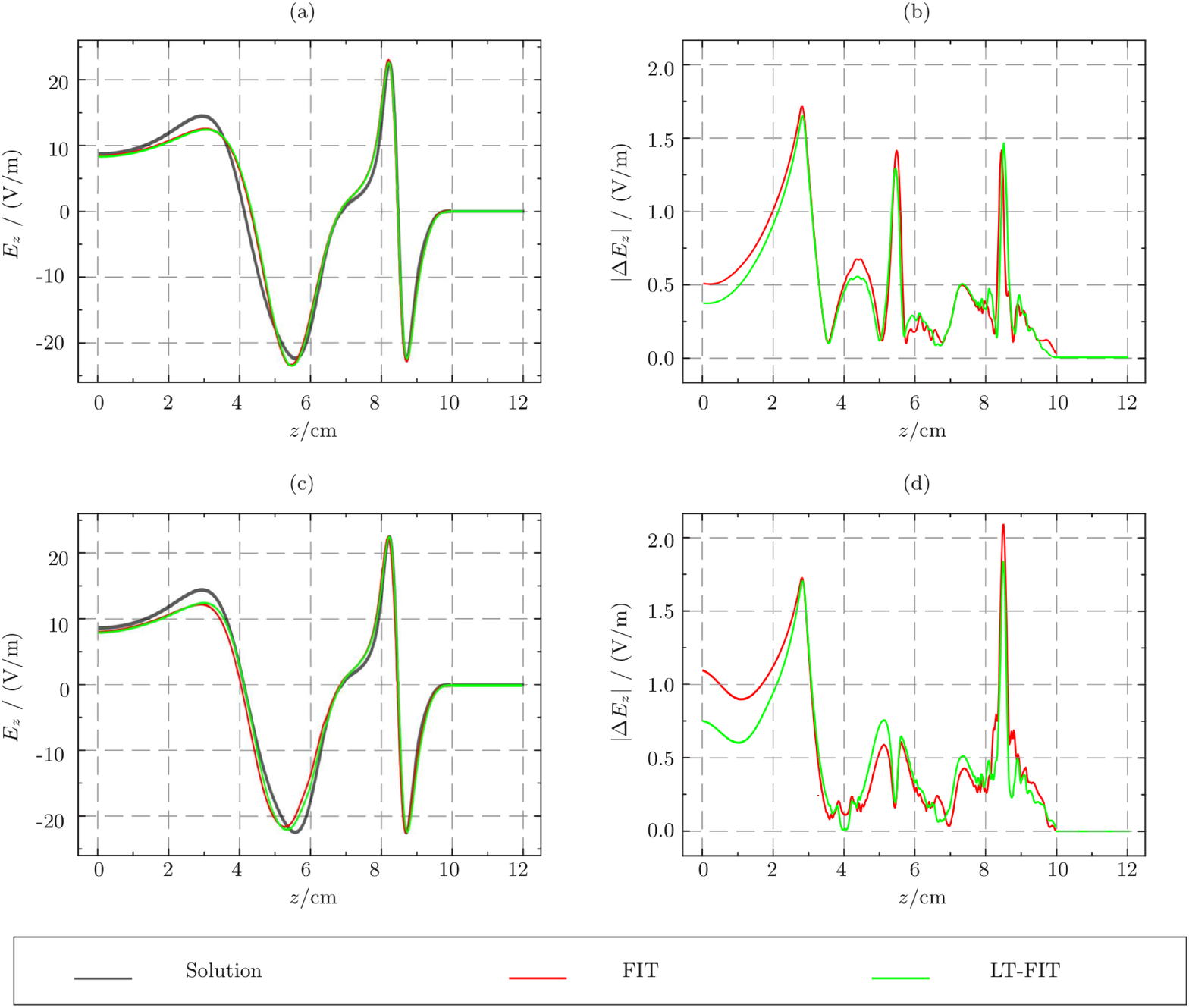}
  \caption{Longitudinal component of the electric field, excited by a {Gaussian} bunch traveling in a pipe.
    The black curve in (a) and (c) indicates the analytical solution for the setup described in Table~\ref{tab:tubeSettings}.
    All other curves shows results obtained by the numerical simulation of the problem. In the top row,
    a static, equidistant computational grid was applied. The results are shown in (a), their errors in (b). In the
    bottom row, time-adaptive grid refinement ($L=3$) was applied. In (c) the results for this case are shown
    and the errors in (d).
  }
  \label{fig:Tube}
\end{figure*}

In the simulations the axis of the cylinder is aligned with the $z$-coordinate.
In the Figs.~\ref{fig:Tube}(a) and \ref{fig:Tube}(c), the analytical solution
of the longitudinal electric field along the axis is depicted by a black curve. The
solutions obtained with the FIT and the LT-FIT are shown in red
and green, respectively.
For the results shown in Fig.~\ref{fig:Tube}(a), an equidistant grid and the
respective maximum time step was applied. In Fig.~\ref{fig:Tube}(b) the errors, given
by the pointwise Euclidean distance of the analytical and the respective numerical solution, are plotted.
The results obtained using dynamic mesh refinement ($L=3$) in the bunch region are
shown in Fig.~\ref{fig:Tube}(c) and the errors in Fig.~\ref{fig:Tube}(d).

A series of simulations of this test setup using various mesh and mesh refinement settings
has been carried out. The results are presented in the Tables~\ref{tab:tubeFIT} (FIT)
and \ref{tab:tubeLTFIT} (LT-FIT).
Besides the mesh settings, they list the number of degrees of freedom (DoF)
applied as well as the computation time in seconds.
The relative $L^2$-error and the \textit{total variation} (TV) of the
longitudinal electric field along the cylinder axis are given
as measures for the quality of the numerical solutions.
The relative $L^2$-error is computed as
\begin{equation*}
  \label{eq:benchRelError}
  \mathcal{E}^\text{rel} = \frac{ \| \mathbf{E}_z - \mathbf{e}_z \|_2}{ \|\mathbf{E}_z\|_2 },
\end{equation*}
with $\mathbf{E}_z = \big(E_z(z(1)), .. E_z(z(N_z)) \big)\transp$
and $E_z$ the longitudinal component of the analytical solution of the electric field
evaluated at the grid point positions $z(i_z)$.

The total variation is a measure for the smoothness of a function~\cite{Harten1983}.
It is defined as
\begin{equation}
  \label{eq:totalVarCont}
  TV(E(z)) := \int_{z(1)}^{z(N_z)} \left| E^\prime(z) \right| \dd z.
\end{equation}
In~\cite{Laney1998} it is shown that the TV of a function equals
the sum of local minima and maxima, where the values at the integration
endpoints count once and all other extrema count twice.
The TV is, therefore, a direct measure for oscillations and their amplitudes.
For discrete solutions it is computed as
\begin{equation}
  \label{eq:totalVarDiscrete}
  TV(\mathbf{e}_z) = \sum\limits_{i_z=1}^{N_z-1} \left| e_z(i_z+1) - e_z(i_z) \right|.
\end{equation}

\begin{table*}[tb]
  \caption{\label{tab:tubeFIT}Results of FIT simulations: the upper half shows results for static, equidistant grids, in the
  lower half the results obtained on time-adaptive grids using Akima splines are displayed.}
  \begin{ruledtabular}
    \begin{tabular}{r|rr|rr|rr|rr}
$L$ & $N_x$ & $N_z$ & $\Delta x$ / mm & $\Delta z_\text{min}$ / mm & DoF / 1e6 & Time / sec & $\mathcal{E}^\text{rel}$ & TV \\
\hline
\multicolumn{4}{l}{} & \multicolumn{2}{c}{static grid} & \multicolumn{3}{l}{}  \\
\hline
0   & 135   & 210   & 0.59            & 0.57                       & 22.96     & 2066       & 0.041              & 1.00  \\
\hline
\multicolumn{4}{l}{} & \multicolumn{2}{c}{time-adaptively refined grid} & \multicolumn{3}{l}{}  \\
\hline
1   & 135   & 105   & 0.59            & 0.57                       & 13.00     & 1225       & 0.044              & 1.42  \\
2   & 135   & 52    & 0.59            & 0.58                       &  9.00     & 940        & 0.040              & 1.93  \\
3   & 135   & 27    & 0.59            & 0.56                       &  8.50     & 925        & 0.041              & 1.99  \\
4   & 135   & 13    & 0.59            & 0.58                       & 11.50     & 1025       & 0.039              & 1.55  \\
    \end{tabular}
  \end{ruledtabular}
\end{table*}
\begin{table*}[tb]
  \caption{\label{tab:tubeLTFIT}Results of LT-FIT simulations: the upper half shows
    results for static, equidistant grids, in the
    lower half the results obtained on time-adaptive grids using Akima splines are displayed.}
  \begin{ruledtabular}
    \begin{tabular}{r|rr|rr|rr|rr}
$L$ & $N_x$ & $N_z$ & $\Delta x$ / mm & $\Delta z_\text{min}$ / mm & DoF / 1e6 & Time / sec & $\mathcal{E}^\text{rel}$ & TV \\
\hline
\multicolumn{4}{l}{} & \multicolumn{2}{c}{static grid} & \multicolumn{3}{l}{}  \\
\hline
0   & 135   & 210   & 0.59            & 0.57                       & 22.96     & 3543       & 0.038              & 0.63  \\
\hline
\multicolumn{4}{l}{} & \multicolumn{2}{c}{time-adaptively refined grid} & \multicolumn{3}{l}{}  \\
\hline
1   & 135   & 105   & 0.59            & 0.57                       & 13.00     & 2008       & 0.045              & 1.39  \\
2   & 135   & 52    & 0.59            & 0.58                       &  9.00     & 1443       & 0.041              & 1.65  \\
3   & 135   & 27    & 0.59            & 0.56                       &  8.50     & 1447       & 0.040              & 1.80  \\
4   & 135   & 13    & 0.59            & 0.58                       & 11.50     & 1791       & 0.037              & 1.40  \\
    \end{tabular}
  \end{ruledtabular}
\end{table*}

The standard FIT is a very well established method for performing PIC simulations.
All TV values are, hence, normalized to the value obtained with the FIT on the finest, nonadaptive
grid, which has been employed.
For comparability, the settings were chosen such that all simulations finish
within less than one hour.

A comparison of the results shows that:
\begin{itemize}
\item[-] The relative errors of the FIT and the LT-FIT are similar, however, the TV of the
LT-FIT solution is significantly lower, indicating reduced oscillations due to
better numerical dispersion properties.
\item[-] For identical numbers of DoF
the LT-FIT simulations take longer because of the higher computational costs for the time integration.
\item[-] For identical mesh resolutions in the bunch area the adaptive simulations yield errors
similar to those obtained on nonadaptive meshes. This follows from a comparison of the result
of the adaptive simulations with the respective nonadaptive simulation.
Hence, the accuracy of the solutions for this example is
mainly determined by the resolution in the bunch region.
\end{itemize}


\subsection{Self-consistent simulation of the PITZ RF gun}
\label{sec:pitz}

In order to drive a free-electron-laser (FEL) operating by the 
self-amplified spontaneous-emission (SASE) principle,
highly charged electron beams of high brightness are required~\cite{Huang2007}.
The aim of the PITZ project (\underline{P}hoto \underline{I}njector \underline{T}est
Facility at DESY \underline{Z}euthen)~\cite{Stephan2004} is the development and
testing of an injector capable of delivering such high quality beams. 
The injectors for the \textit{\underline{F}ree Electron \underline{Las}er 
in \underline{H}amburg} (FLASH)~\cite{Ackermann2007:long} and the future \textit{European X-Ray Laser Project}
XFEL~\cite{XFEL2} are under development at PITZ.

\begin{figure*}[htb]
    \psfrag{cat}[lc]{photo cathode}
    \psfrag{cav}[lc]{resonant cavity}
    \psfrag{ant}[lc]{coupling antenna}
    \psfrag{cou}[lc]{RF input coupler}
    \psfrag{path}[lc]{electron path}
    \psfrag{valv}[lc]{shutter valve}
    \psfrag{cross}[lc]{diagnostics section}
    \psfrag{mirror}[lc]{laser mirror}
    \psfrag{pump}[lt]{to vacuum pump}
    \includegraphics[width=\textwidth]{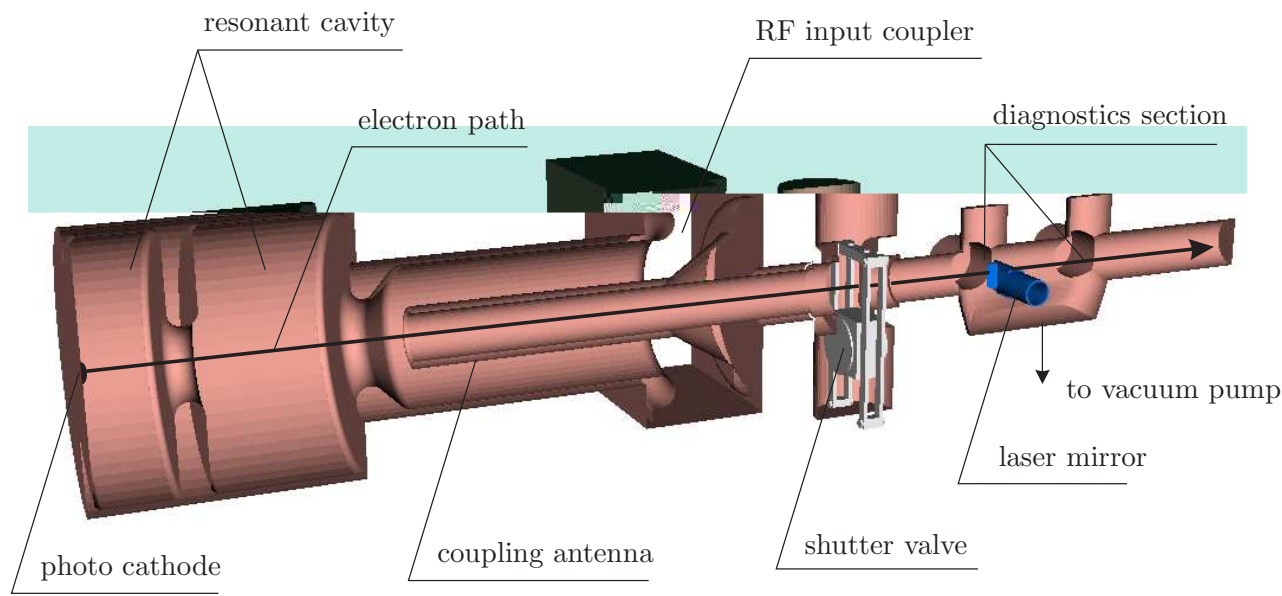}
  \caption{A CAD model of the injector section of the \textit{Free-electron LASer in Hamburg} (FLASH)
      is shown in cut view. The depicted part corresponds to approximately 60\cm\ of the injector. In total 2~m
      have been modeled. However, the non-depicted part consists only of  the beam pipe.
      A short laser pulse is directed onto the photo cathode, where electrons are emitted.
      The bunch of electrons is accelerated by a high-frequency
      electromagnetic field, which is excited in the resonant cavity. It propagates
      along the electron path through the coupling antenna and passes the shutter valve and
      the laser mirror, which is inserted through one arm of
      the diagnostics section. The other arms
      are used for inserting measurement devices. The opening on its bottom side
      is connected to a vacuum pump.
  }
  \label{fig:PITZgun3D}
\end{figure*}

The layout of the radio-frequency (RF) gun of the injector is shown in Fig.~\ref{fig:PITZgun3D}.
The emitted bunch is accelerated
in a 1.5-cell L-band cavity providing for an accelerating gradient of 42~MV/m at
an operational frequency of 1.3~GHz.
A focusing technique proposed in~\cite{Carlston1989} is applied in order to
compensate for the correlated part of the space charge induced emittance growth. 
The initial gun layout was designed such that a minimum of the transverse emittance is
expected at a distance of approximately 1.6~m downstream of the cathode. At this position
a RF cavity is installed in order to accelerate the electrons
to relativistic energies. The main design parameters are listed in Table~\ref{tab:PITZparameters}.


\begin{table}[htb]
  \caption{\label{tab:PITZparameters}PITZ design parameters}
  \begin{ruledtabular}
    \begin{tabular}{lr}
    Parameter                  & Design value \\
    \hline
    Bunch charge               & 1 nC\\
    Transverse laser profile   & Hat-profile\\
    Laser spot radius          & 1 mm\\
    Longitudinal laser profile & Flat-top\\
    Laser pulse duration       & 22 ps\\
    Rise/Fall time             & 2 ps\\
    Accelerating gradient      & 42 MV/m\\
    Transverse emittance       & 1 mm mrad\\
    \end{tabular}
  \end{ruledtabular}
\end{table}

Many numerical studies of the injector using PIC codes have been performed over the last years.
Results have been published, \eg in
\cite{Zhang1996,Cee2002a,Setzer2004,Schnepp2006c,Schnepp2007a}.
However, these simulations either assumed a rotationally symmetric geometry and made
use of a so-called 2.5-dimensional approach
as implemented in the MAFIA TS2 module~\cite{CST}, or otherwise very short distances
in the cm-range of the full
three-dimensional model were simulated. Results of parallelized simulations
of the full three-dimensional model up to one meter
downstream of the cathode
have been presented in~\cite{Wolfheimer2006,Gjonaj2006}. However,
the exact position of the transverse emittance minimum is required
for the optimal placement of the accelerating cavity. Also, the
value of the emittance at this position is of strong interest in order
to check whether the design value is met.

In the context of this work,
the gun was simulated up to a distance of two meters downstream of the cathode.
First, settings of the simulation parameter such as the grid resolution and the number
of macro particles have been determined.
Then, a design study was performed, which addresses the effects of individual
injector elements on the beam quality.

\begin{figure*}[htb]
    \psfrag{25}[cc]{\small 2.5}
    \psfrag{20}[cc]{\small 2.0}
    \psfrag{15}[cc]{\small 1.5}
    \psfrag{10}[cc]{\small 1.0}
    \psfrag{05}[cc]{\small 0.5}
    \psfrag{12}[cc]{\small 1.2}
    \psfrag{14}[cc]{\small 1.4}
    \psfrag{16}[cc]{\small 1.6}
    \psfrag{18}[cc]{\small 1.8}
    \psfrag{19}[cc]{\small 1.9}
    \psfrag{20}[cc]{\small 2.0}
    \psfrag{21}[cc]{\small 2.1}
    \psfrag{22}[cc]{\small 2.2}
    \psfrag{23}[cc]{\small 2.3}
    \psfrag{RZ}[cb]{$\sigma_z$ / mm}
    \psfrag{z}[cc]{$z$ / cm}
    \psfrag{a}[cc]{(a)}
    \psfrag{c}[cc]{(b)}
    \includegraphics[width=0.99\textwidth]{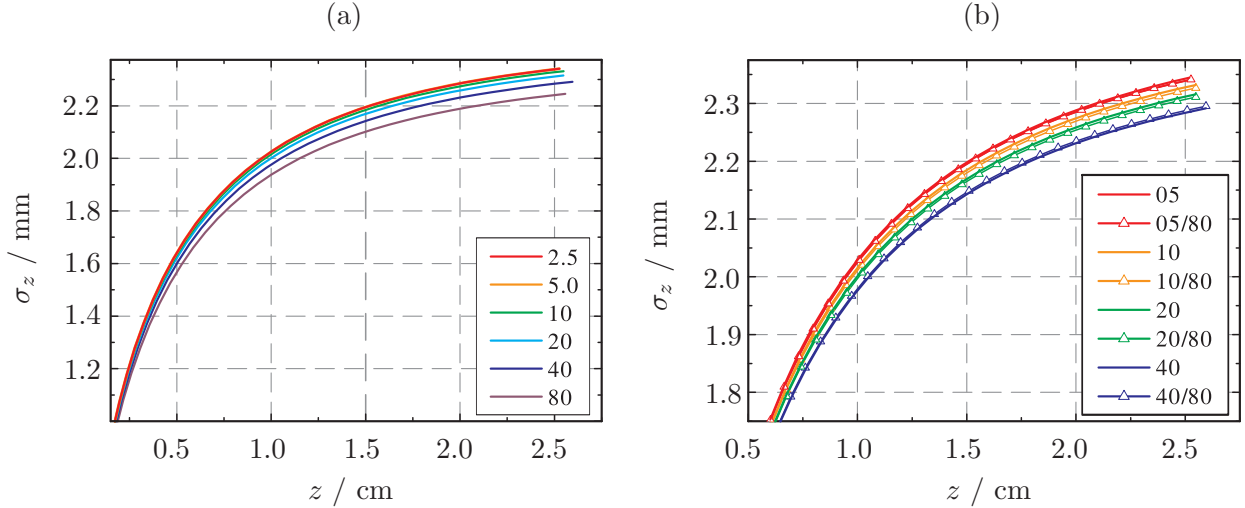}
  \caption[Computed RMS bunch length vs.~longitudinal grid step size]{
    Computed RMS bunch length vs.~longitudinal grid step size.
    In (a) equidistant grids with different grid step sizes were used.
    In (b) the equidistant grid with a step size of 80\um\ was
    statically refined within the first cm from the cathode.
    The step sizes given in the legend are in units of\um.
  }
  \label{fig:ApplPITZConvRMSz}
\end{figure*}

Previous investigations have shown that the length
of the simulated bunches critically depends on the longitudinal
grid step size~\cite{Schnepp2006}. In~\cite{Zhang1996}
it was stated that a grid resolution of 20\um\ longitudinally is required
in order to obtain accurate results.
In the Fig.~\ref{fig:ApplPITZConvRMSz}(a) the results of a 
parameter study addressing the relation of the longitudinal grid step size and the computed bunch length
are presented.
The initial 2.5~cm of the gun were simulated using an equidistant grid
of 2.5\um\ to 80\um\ step size. The computed RMS bunch length $\sigma_z$ at $z = 2.5\cm$
varies from 2.24\mm\ to 2.35\mm.
Changing the step size from 5\um\ to 2.5\um\ results in a variation
of $\sigma_z$ by approximately 1\um.
The computed bunch lengths for a step size of 2.5\um\
and 10\um\ differ by approximately 10\um. 
A longitudinal grid step of 10\um\ is considered
to be a reasonable balance of computational costs and accuracy.

After emission the electrons have a very low energy of $\approx5$\eV\
and space charge forces have a strong influence. Since the particles gain energy quickly, it
is sufficient to apply the smallest grid step size only in the immediate vicinity of the cathode.
Hence, an additional static grid refinement is applied in this area.
In the Fig.~\ref{fig:ApplPITZConvRMSz}(b) results for this approach are shown. 
A grid with a uniform step size of 80\um\ is statically refined within the first centimeter from
the cathode. Refinement levels from one to four have been applied,
resulting in step sizes of 40, 20, 10, and 5\um\ within the refined region.
The results are, except for minor discrepancies,
identical to those obtained with an equidistant
grid of the same minimum step size.

For the simulation of the full structure,
a combination of static and dynamic mesh refinement
was applied providing for a longitudinal resolution of 10\um\ for the first cm from the cathode
and 80\um\ thereafter. The requirements on the transverse grid step size are less
demanding. The width of the simulated bunches using a step size of 2.5\um\
and 80\um\ differ by less than 5\um. Deviations
in the\um-range correspond to some per mille of the bunch width.
Choosing a minimum transverse step size of 80\um\ and a refinement level
of five, we obtained an adaptive computational mesh consisting of an average
of approximately 82 million cells. A nonadaptive mesh providing for the
same minimum resolution would consist of approximately 1.7 billion cells.
The bunches in the simulations comprised approximately 0.5 million
macro particles.





\begin{figure*}[htb]
    \centering
    \psfrag{z}[cc]{$z$ / m}
    \psfrag{a}[cc]{(a)}
    \psfrag{c}[cc]{(b)}
    \psfrag{e}[cc]{(c)}
    \psfrag{n}[cc]{(d)}
    \psfrag{o}[cc]{(e)}
    \psfrag{r}[cc]{(f)}
    \small
    \psfrag{00}[cc]{0.0}
    \psfrag{05}[cc]{0.5}
    \psfrag{10}[cc]{1.0}
    \psfrag{15}[cc]{1.5}
    \psfrag{20}[cc]{2.0}
    \psfrag{22}[cc]{2.2}
    \psfrag{24}[cc]{2.4}
    \psfrag{26}[cc]{2.6}
    \psfrag{25}[cc]{2.5}
    \psfrag{30}[cc]{3.0}
    \psfrag{25}[cc]{2.5}
    \psfrag{40}[cc]{4.0}
    \psfrag{35}[cc]{3.5}
    \psfrag{13}[cc]{1.3}
    \psfrag{14}[cc]{1.4}
    \psfrag{16}[cc]{1.6}
    \psfrag{17}[cc]{1.7}
    \psfrag{18}[cc]{1.8}
    \psfrag{19}[cc]{1.9}
    \psfrag{600}[rc]{600}
    \psfrag{400}[rc]{400}
    \psfrag{200}[rc]{200}
    \psfrag{000}[rc]{0}
    \normalsize
    \psfrag{RX}[cc]{$\sigma_x$ / mm}
    \psfrag{RZ}[cc]{$\sigma_z$ / mm}
    \psfrag{EN}[cc]{Energy / MeV}
    \psfrag{EX}[cc]{$\emit_x$ / (mm mrad)}
    \includegraphics[width=0.99\textwidth]{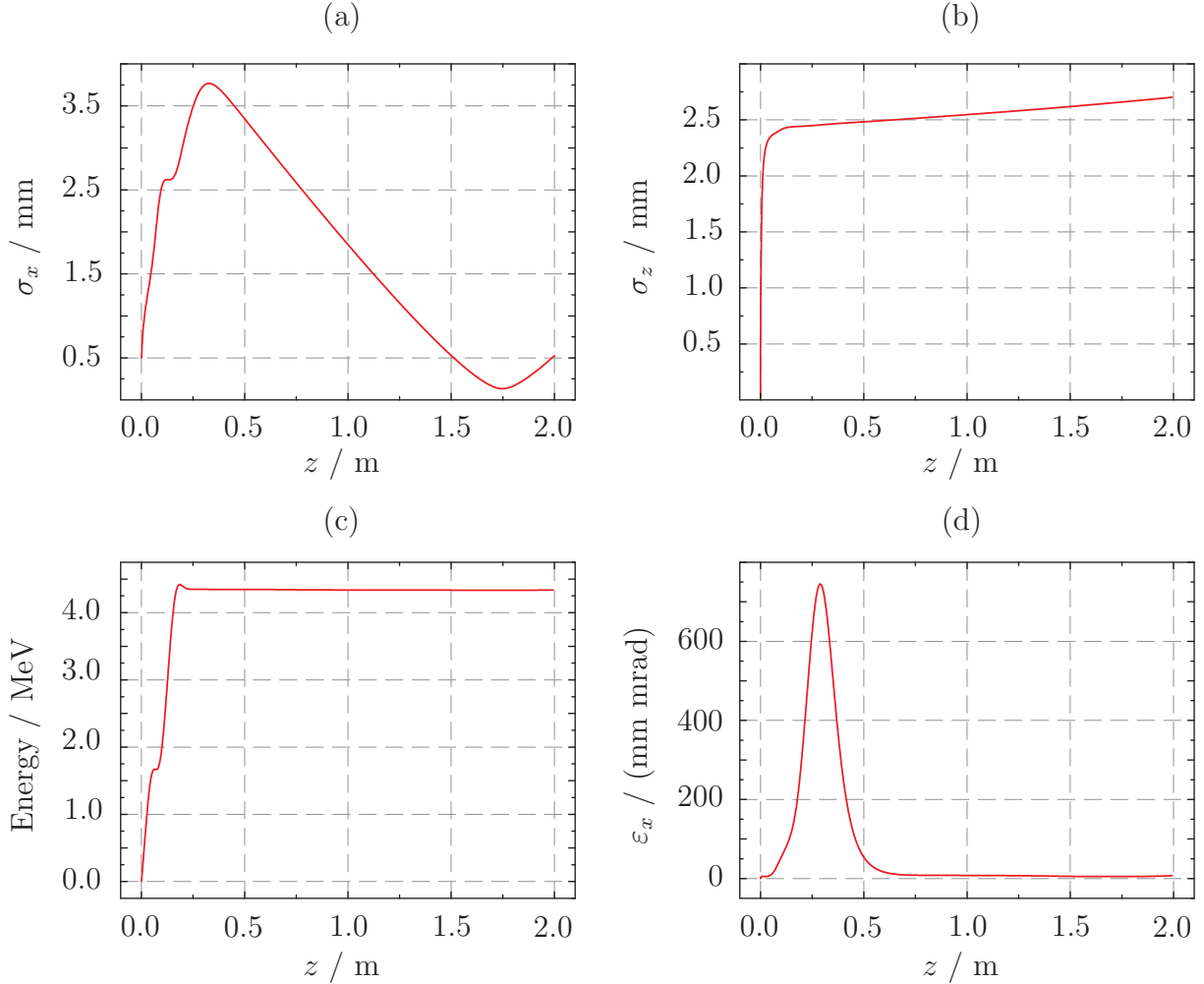}
  \caption[Evolution of bunch parameters in the PITZ RF gun]{
    The plots show the evolution of the RMS bunch width (a), RMS bunch length (b),
    average particle energy (c), and horizontal emittance (d) along
    the longitudinal coordinate of the PITZ RF gun.
    \label{fig:PITZresults}}
\end{figure*}

The results given in the following were obtained 
with the LT-FIT method. In Fig.~\ref{fig:PITZresults}, the evolution
of the horizontal root-mean-square (RMS) bunch width $\sigma_x$ (a),
the RMS bunch length $\sigma_z$ (b), the average particle energy (c),
and the horizontal emittance $\emit_x$ (d) for the model shown
in Fig.~\ref{fig:PITZgun3D} are plotted.
The projected emittances $\emit_x$ and $\emit_y$ are given by
$  \emit_u  = \left( \sigma^2_u \cdot \sigma^2_{p_u} - (\sigma_{u,p_u})^2 \right)^{\nicefrac{1}{2}}, 
  u \in \{x,y\},$
where $\sigma^2_{p_u}$ is the momentum variance and $\sigma_{u,p_u}$ is the covariance
of position and momentum.  

\begin{figure*}[htb]
    \centering
    \psfrag{z}[cc]{$z$ / m}
    \psfrag{a}[cc]{(a)}
    \psfrag{c}[cc]{(b)}
    \psfrag{e}[cc]{(c)}
    \psfrag{n}[cc]{(d)}
    \psfrag{o}[cc]{(e)}
    \psfrag{r}[cc]{(f)}
    \small
    \psfrag{15}[cc]{1.5}
    \psfrag{20}[cc]{2.0}
    \psfrag{22}[cc]{2.2}
    \psfrag{24}[cc]{2.4}
    \psfrag{26}[cc]{2.6}
    \psfrag{16}[cc]{1.6}
    \psfrag{17}[cc]{1.7}
    \psfrag{18}[cc]{1.8}
    \psfrag{19}[cc]{1.9}
    \normalsize
    \psfrag{RX}[cc]{$\sigma_x$ / mm}
    \psfrag{RZ}[cc]{$\sigma_z$ / mm}
    \psfrag{EN}[cc]{Energy / MeV}
    \psfrag{EX}[cc]{$\emit_x$ / (mm mrad)}
    \psfrag{EY}[cc]{$\emit_y$ / (mm mrad)}
    \psfrag{C}[lc]{Gun}
    \psfrag{D}[lc]{+ Diag. Cross}
    \psfrag{E}[lc]{+ Mirror}
    \psfrag{F}[lc]{+ Valve}
    \includegraphics[width=0.99\textwidth]{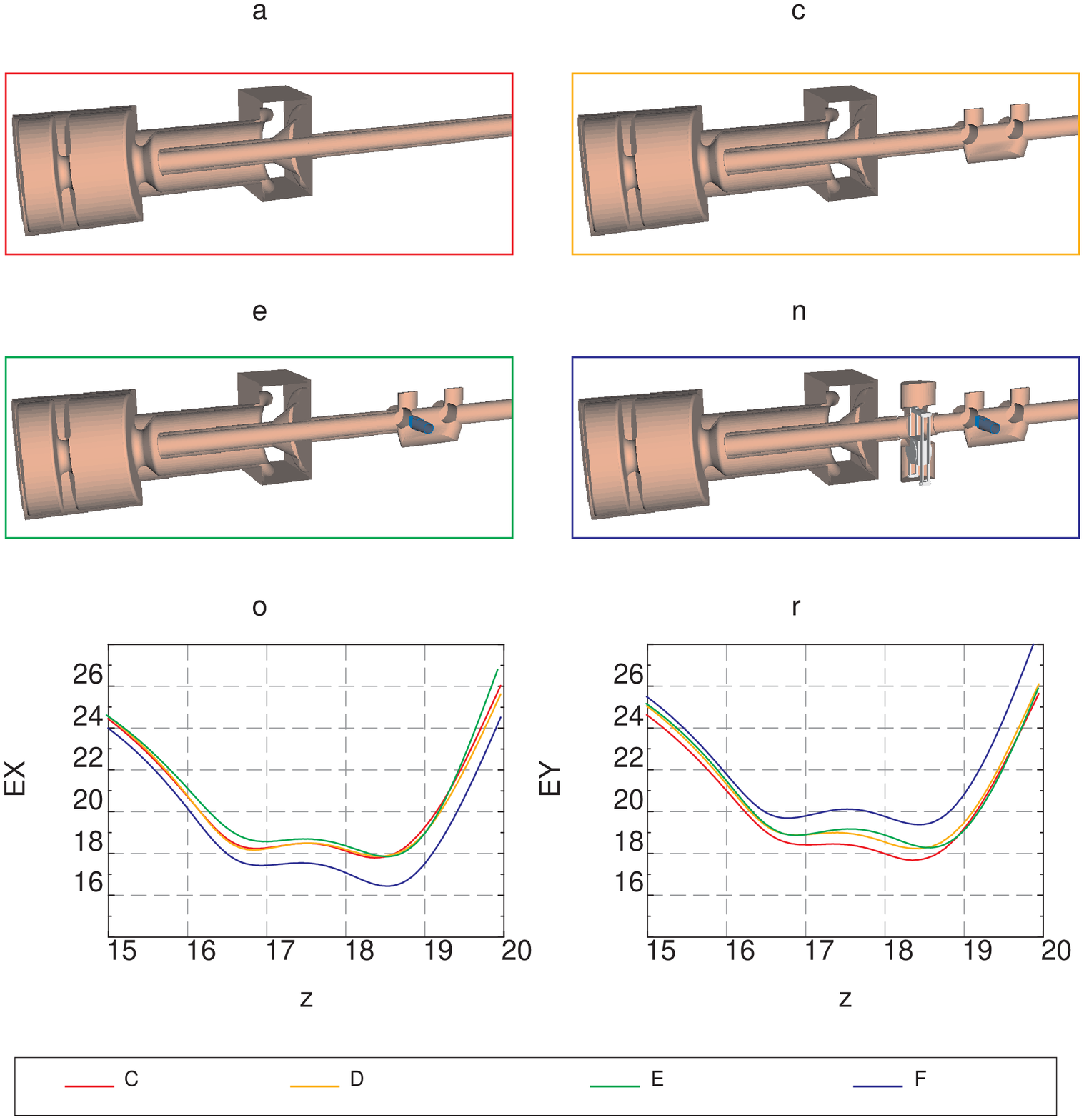}
  \caption[Models and results of the PITZ gun design study]{
    Models and results of the PITZ RF gun design study.
    In (a) to (d) the CAD models of the simulated structures are shown. 
    The horizontal and vertical emittances around the position of
    their minimum are plotted in (e) and (f) respectively. The colors
    of the curves correspond to the outline color of the
    models above. While $\emit_x$ and $\emit_y$ are similar for
    the case (a), they show a distinct asymmetry for the model depicted in (d).
  }
  \label{fig:PITZemit}
\end{figure*}

We performed a design study of the injector in order
to identify the individual effects of the diagnostics section, the laser mirror, and
the shutter valve on the beam quality in terms of emittance growth. The respective
models are shown in the Figs.~\ref{fig:PITZemit}(a)-(d). For each model, an additional element
is added. The colors of the curves in Fig.~\ref{fig:PITZemit}(e)-(f)
correspond to the outline color of the respective model.

For the model given in Fig.~\ref{fig:PITZemit}(a), the horizontal and vertical emittances, $\emit_x$ and $\emit_y$,
are identical. This is expected since the geometry is rotationally symmetric
for all parts close to the beam. The RF input coupler is obviously not rotationally
symmetric but this symmetry violation is hidden by the coupling antenna.
In Fig.~\ref{fig:PITZemit}(b), the large opening on the bottom side of the doublecross introduces an
asymmetry of the geometry in horizontal and vertical direction. While $\emit_x$ is almost unchanged,
the value of $\emit_y$ increases by approximately 0.04\mm~\milli rad. The laser mirror,
added in the horizontal plane, introduces another asymmetry in Fig.~\ref{fig:PITZemit}(c). An influence
on the transverse emittances can be identified in the results but
the actual emittance growth at the position of the minimum is small.
Finally, the shutter valve depicted in Fig.~\ref{fig:PITZemit}(d) results in significant
changes of the transverse emittances. While the horizontal emittance actually decreases
by approximately 0.1\mm~\milli rad, the vertical emittance increases by about
0.18\mm~\milli rad in comparison to the structure of Fig.~\ref{fig:PITZemit}(a). Hence, the shutter valve
causes a distinct asymmetry in the transverse beam dynamics.
In the meantime, a shutter valve with RF shieldings has been installed.



%




\section{Conclusions}
\label{sec:conclusions}

In this paper, we extended the framework of Finite Integration Technique
to include dynamic mesh refinement. This offers computability
on workstations for a class of multi-scale problems, which before was accessible
to massively parallelized computations only. In particular, our approach
enabled us to perform
fully self-consistent simulations of the PITZ RF gun up to two meters downstream from
the cathode. We elaborated on the details concerning
the mesh refinement and coarsening procedures and presented
a novel type of sub-spline interpolation, which is entirely devoid of overshooting.
Finally, we performed a design study, which identifies
the emittance growth due to individual parts of the gun.

\begin{acknowledgments}
The work of S.~Schnepp is supported by the 'Initiative for Excellence' of the German 
Federal and State Governments and the Graduate School of Computational 
Engineering at Technische Universit\"at Darmstadt. The authors thank Thomas Lau for
providing the analytical solution to the example used in Sec.~\ref{sec:pipe}.
\end{acknowledgments}

%

\end{document}